\documentclass[5pt,final]{iopart}
\expandafter\let\csname equation*\endcsname\relax
\expandafter\let\csname endequation*\endcsname\relax
\usepackage{amsmath}

\usepackage{amssymb}

\usepackage[makeroom]{cancel}

\usepackage{mhchem}

\usepackage{graphicx}

\usepackage{dcolumn}

\usepackage{bm}
\usepackage[normalem]{ulem}
\usepackage{color}

\usepackage{verbatim}

\usepackage{epstopdf}

\newcommand{\bbr}{\mathbf{r}}

\newcommand{\sep}{ \ \ \ , \ \ \ }

\newcommand{\beq}{\begin{equation}}
\newcommand{\eeq}{\end{equation}}
\newcommand{\beqn}{\begin{eqnarray}}
\newcommand{\eeqn}{\end{eqnarray}}
\newcommand{\pp}{\partial}
\newcommand{\dd}{{\rm d}}
\newcommand{\ee}{{\rm e}}
\newcommand{\eq}{Eq.\ }

\newcommand{\eqs}{Eqs }
\newcommand{\fig}{Fig.\ }
\newcommand{\figs}{Figs~}

\newcommand{\hatPin}{\hat{P}_{\rm in}}
\newcommand{\hatPo}{\hat{P}_{\rm out}}

\newcommand{\cred}{\color{red}}

\begin{document}

\title[Stress granule formation via phase separation]{Stress granule formation via ATP depletion-triggered phase separation
}

\author{Jean David Wurtz}
\address{Department of Bioengineering, Imperial College London, South Kensington Campus, London SW7 2AZ, U.K.}
\author{Chiu Fan Lee}
\address{Department of Bioengineering, Imperial College London, South Kensington Campus, London SW7 2AZ, U.K.}
\ead{c.lee@imperial.ac.uk}
\begin{abstract}
Stress granules (SG) are droplets of proteins and RNA 
that form in the cell cytoplasm during stress conditions. We consider minimal models of stress granule formation based on the mechanism of phase separation regulated by ATP-driven chemical reactions. Motivated by experimental observations, we identify a minimal model of SG formation triggered by ATP depletion. Our analysis indicates that ATP is continuously hydrolysed to deter SG formation under normal conditions, and we provide specific predictions that can be tested experimentally.
\end{abstract}
\submitto{\NJP}
\maketitle

\section{Introduction} 
\label{sec:NJP:intro}
In order to function, a biological cell has to organise its interior contents into compartments that can dynamically evolve depending on the environment and  moment in the cell cycle. For example, cytoplasmic stress granules ({\bf SG}) form when the cell is under stress \cite{protter_trends16}
  and P granules localise to one side of the embryo of {\it C. elegans}  prior to the first cell division \cite{updike_andrology10,brangwynne_science09,lee_prl13}.
   Both SG and P granules belong to the class of  ribonucleoprotein  ({\bf RNP}) granules. RNP granules consist of RNA and proteins and lack a surrounding membrane. Recent experiments provided strong evidence that 
the intriguing mechanism behind their formation and dynamical regulation, in the absence of a membrane, is  based on the physical phenomenon of phase separation  \cite{hyman_annrev14, brangwynne_natphys15}. Phase separation refers to the spontaneous partitioning of a system into compartments of distinct macroscopic properties, such as the condensation of water vapour into droplets when fog forms. Phase separation under equilibrium condition is a well-understood phenomenon \cite{bray_advphys02}, however the phase separation in the cytoplasm can be driven out of equilibrium by a number of effects that include activity of motor proteins shuttling along the cytoskeleton \cite{ivanov_expcell03}, and chemical reactions affecting the phase-separating behaviour of RNP granule constituents \cite{zwicker_pnas14,zwicker_pre15,saha_cell16,weber_njp17,wurtz_prl18}. 

In this paper, we  apply the theoretical formalism developed in \cite{wurtz_prl18} to investigate minimal models of SG regulation based on non-equilibrium, chemical reaction-controlled phase separation. SG are a particularly dynamic type of  RNP granules as they form quickly, in the order of 10 min, 
 when the cell is under stress (e.g., heat shock, chemical stress, osmotic shock, etc.), and also dissolve away rapidly when the stress is removed \cite{wheeler_elife16, ohshima_plos15}.  The cell's reaction to external stresses by forming SG  is important for its survival \cite{anderson_currbiol09,protter_trends16}. Although specific functions of SG remain unclear, they are thought to be involved in protecting messenger-RNA
by recruiting them into SG away from harmful conditions \cite{aulas_jcb15, buchan_cell16, kedersha_jcb05}. In addition, SG malfunction is associated  with several degenerative diseases such as amyotrophic lateral sclerosis  and multisystem
proteinopathy \cite{ramaswami_cell13}.

%
%

\subsection{Experimental observations}
\label{sec:NJP:experiments}

Experimental studies have shown that  SG assemble in response to multiple types of stress situations \cite{hofmann_mboc12, kedersha_jcb99, kedersha_mboc02}, and several pathways for SG formation have been identified. The most established is the ATP-dependent phosphorylation of the translation initiation factor eIF2$\alpha$ causing the arrest of RNA translation \cite{anderson_currbiol09}. RNA is subsequently released from polysomes and aggregate with various proteins to form SG. There exist also other pathways that are independent of eIF2$\alpha$ phosphorylation 
	 \cite{dang_jbc06}, such as energy starvation {\cite{kedersha_mboc02}}. { Indeed}, while various stress conditions causing SG assembly also cause a depletion of the cytoplasmic energy stores \cite{hofmann_mboc12, lilly_cancerres84, chang_ajp01}, energy starvation alone has been shown to trigger SG formation \cite{kedersha_mboc02}. Here, we will focus exclusively on how changes in the energy level, and specifically the ATP concentration, can regulate SG formation. In other words, we will make the assumption that the ATP concentration directly triggers SG formation through ATP-dependent biochemical reactions.

%
In addition to this central premise, 
we will use two other pieces of biological observations to guide our modelling: 
 First, during normal conditions, i.e. without imposed stress, the ATP concentration is at the normal level and SG are absent, or at least are so small that they are undetectable microscopically. Second, when external stress is imposed, ATP can fall by 50\% \cite{hofmann_mboc12, chang_ajp01}, in a time scale similar to that of SG formation \cite{hofmann_mboc12, lilly_cancerres84, chang_ajp01},  and SG assemble with sizes of the order of a micrometer \cite{anderson_currbiol09}. 
 We will therefore impose the following two constraints on our modelling: 1) {\it a fall in the ATP level leads to the growth of SG}, and 2) {\it the response to the ATP the level is switch-like,} in the sense that a relatively mild change in the ATP level can lead  to a very large change in SG size.
	
We now set out to construct minimal models that are compatible with these salient experimental findings on SG regulation. 

\begin{figure}
	\centering
	\includegraphics[scale=1]{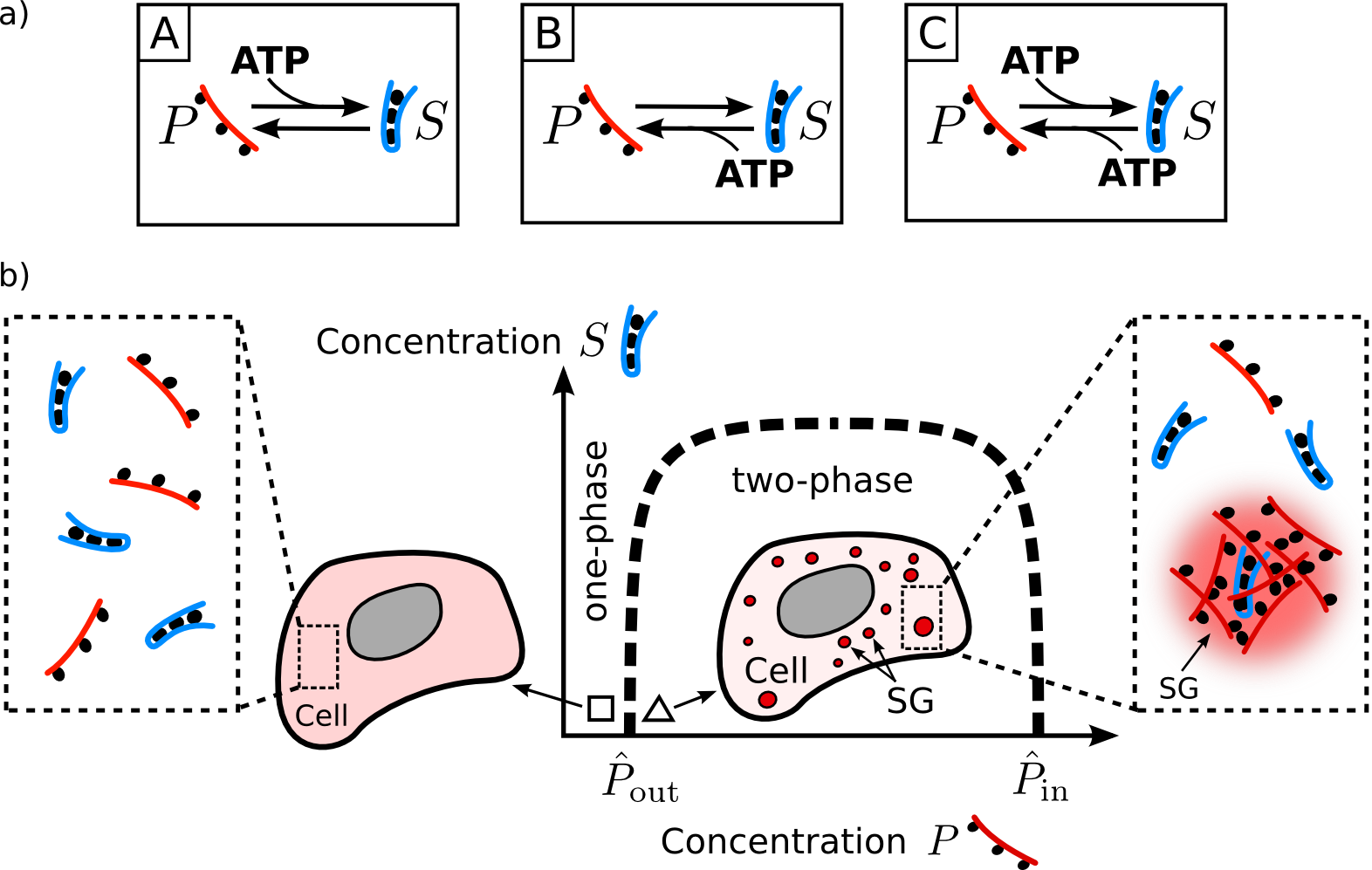}
	\caption{(a) {\it Distinct schemes of ATP-driven chemical reactions that control cytoplasmic SG formation.} The phase-separating form of the constituent molecules in SG is denoted by $P$ and the soluble form is denoted by $S$. In model A, ATP promotes conversion from $P$ to $S$, while it is the reverse in model $B$. In model $C$, ATP drives both conversions.
(b)	\textit{Phase diagram.}
		At low $P$ concentration, molecules distribute homogeneously (`$\square$' symbol). When the concentration of $P$ increases beyond the phase boundary denoted by the  dashed line, $P$-rich droplets form and are surrounded by a $P$-poor  
		phase (`$\triangle$' symbol).
	}  \label{fig:NJP:fig1}
\end{figure}
\section{Minimal models}
\label{sec:NJP:models}
We first describe a set of minimal models of SG regulation based on the principle of  phase separation controlled by ATP-driven chemical reactions. 
A similar reaction scheme has recently been employed to study centrosomes \cite{zwicker_pnas14}, and the associated physics has started to be explored in \cite{zwicker_pre15,wurtz_prl18}.
We consider the cell cytoplasm as a ternary mixture of molecules: the phase-separating form of the SG constituent molecules ($P$), the soluble form of the same molecules ($S$), and the remaining molecules in the cytoplasm ($C$). Note that the constituents of SG 
consist of many distinct proteins and mRNA \cite{protter_trends16}. Therefore, $P$ and $S$ are meant to represent the average behaviour of the set of proteins and mRNA responsible for SG formation via phase separation \cite{jacobs_biophysj17}. The same applies to the component $C$, which represents the average behaviour of the cytoplasmic molecules not involved in SG formation. To control phase separation, we further assume that $P$ and $S$ can be inter-converted by chemical reactions that are potentially ATP-driven:
\begin{equation}
\label{eq:NJP:reactions}
\ce{
$P$
<=>[$k$][$h$] 
$S$ 
}
\end{equation}
where $k,h$ denote the forward and backward reaction rate constants. In the biological context these reactions can be protein post-transcriptional modifications. For example the phase behaviour (phase-separated or homogeneous) of intrinsically disordered proteins, a class of proteins that lack a well defined secondary structure, can be controlled via their phosphorylation/dephosphorylation \cite{bah_jbc16, li_nature12}. In our minimal description, we assume that there  is no cooperativity in the chemical reactions, i.e. $k,h$ do not depend on the concentrations of the molecular components. However, the rates can be influenced by   the ATP concentration, denoted by $\alpha$, in a linear manner. With these simplifications, we can categorise the distinct schemes into three models: A, B, and C (\fig \ref{fig:NJP:fig1}(a)). In model A, ATP promotes the conversion from the phase-separating state $P$ to the soluble state $S$. 
In model B, ATP promotes the reverse reaction, and in model C, both reactions are driven by ATP. Specifically, we have:
\beqn
\label{eq:models}
\label{eq:NJP:modelA}
{\rm Model \ A:} \quad  \quad k_A(\alpha) &=& \alpha K_A,\quad  h_A(\alpha) =H_A \\
\label{eq:NJP:modelB}
{\rm Model \ B:} \quad \quad k_B(\alpha) &=&  K_B, ~~\quad h_B(\alpha) =\alpha H_B  \\
\label{eq:NJP:modelC}
{\rm Model \ C:} \quad \quad k_C(\alpha) &=& \alpha K_C, \quad h_C(\alpha) =\alpha H_C \ ,
\eeqn
with $K_i,H_i$ being constants, and where $i =A,B,C$ refers to the model under consideration. 
The absence of a constant term in the ATP-driven reaction rates implies that these reactions cannot occur spontaneously (without ATP). This is justified by the fact that the free energy barrier for ATP-driven reaction is typically high, rendering spontaneous reactions negligible \cite{ardito_ijmm17}.

\section{Steady-state of a  multi-droplet system}
\label{sec:NJP:steadyStateAndStability}
Given the three minimal models depicted in \fig \ref{fig:NJP:fig1}(a), we now aim to  determine which one of them is the most compatible with experimental observations. To do so, we will first elucidate the salient features of the mechanism of chemical reaction-driven phase separation  
 by considering a single phase-separated droplet of radius $R$ co-existing with the dilute medium.

We assume that both inside and outside SG, $P$ and $S$ undergo the chemical reactions described by \eqs \eqref{eq:NJP:modelA}-\eqref{eq:NJP:modelC}, and free diffusion with the same diffusion coefficient $D$. This approximation is supported by experimental observations that show that SG are not inert, highly packed molecular aggregates, but porous \cite{souquere_jcs09} and dynamic, with SG constituents rapidly shuttling in and out \cite{kedersha_jcb00,bley_17}. Then, denoting the concentrations of $P$ and $S$ by the same symbols,  they obey the reaction-diffusion equations: 
\begin{subequations} \label{eq:main}
	\begin{align}
		\label{eq:NJP:P}
	\pp_t P_{\rm in/out} &= D\nabla^2 P_{\rm in/out} - k_i(\alpha) P_{\rm in/out} +h_i(\alpha) S_{\rm in/out}
	\\
			\label{eq:NJP:SS}
	\pp_t S_{\rm in/out} &= D\nabla^2 S_{\rm in/out} + k_i(\alpha) P_{\rm in/out} -h_i(\alpha) S_{\rm in/out}
	\ ,
	\end{align}
\end{subequations}
where $i =A,B,C$ and the subscripts ``in'' and ``out'' denote  the concentration profiles inside and outside the droplet. 

In addition to the dynamical equations, we need to supply the boundary conditions. To this end, we  assume further that the system is close to equilibrium to the extent that local thermal equilibrium applies around the droplet interfacial region, which means that 
\beqn
\label{eq:interfacePin}
P_{\rm in}(R_-) &=&\hat{P}_{\rm in} 
\\
\label{eq:NJP:gt}
P_{\rm out}(R_+) &=&\hat{P}_{\rm out} \left(1+\frac{l_c}{R}\right)
\\
\label{eq:NJP:S}
S_{\rm in} (R_-)&=&S_{\rm out} (R_+)
 \ ,
\eeqn
where we have assumed spherical symmetry with the centre of the droplet being the origin of our coordinate system. 
In the equations above,  $R_\pm \equiv R \pm \epsilon$ with $\epsilon$ being an  infinitesimal constant. In other words, $R_+$ denotes the position just outside the droplet and  $R_-$ denotes the position just inside the droplet. 
$\hat{P}_{\rm in/out}$ are the equilibrium coexistence concentrations given by the phase boundary (see \fig \ref{fig:NJP:fig1}(b)), and \eq (\ref{eq:NJP:gt}) is the Gibbs-Thomson relation with $l_c$ being the capillary length. The concentration just outside droplets therefore increases as the droplet radius $R$ decreases, which is a consequence of the Laplace pressure \cite{bray_advphys02}. In addition, we assume that $S$ is inert to phase separation  
so that
 its concentration profile
is continuous at the interface,
 hence \eq (\ref{eq:NJP:S}). 
The boundary conditions above are further complemented by the no-flux boundary conditions:
\beq
\label{eq:NJP:boundary2}
0=\pp_r P_{\rm in}(r)\Big|_{r=0} = \pp_r S_{\rm in}(r)\Big|_{r=0}  = \pp_r P_{\rm in}(r)\Big|_{r=L} = \pp_r S_{\rm in}(r)\Big|_{r=L}
\ ,
\eeq
where $L$ is the system size and $r$ is the distance from the droplet's centre.

Another crucial approximation we will employ is the {\it quasi-static assumption} as done in the Lifshitz-Slyozov theory for equilibrium phase separation \cite{lifshitz_jpcs61}, 
which assumes that the time scale at which concentration profiles reach a steady-state is much faster than the time scale at which a droplet radius changes by a non-negligible amount.
With this separation of time scale, we only need to solve for the steady-state solutions of  \eqs (\ref{eq:main})  in order to study 
the droplet's dynamics in the system.

\begin{figure}
	\centering
	\includegraphics[scale=1]{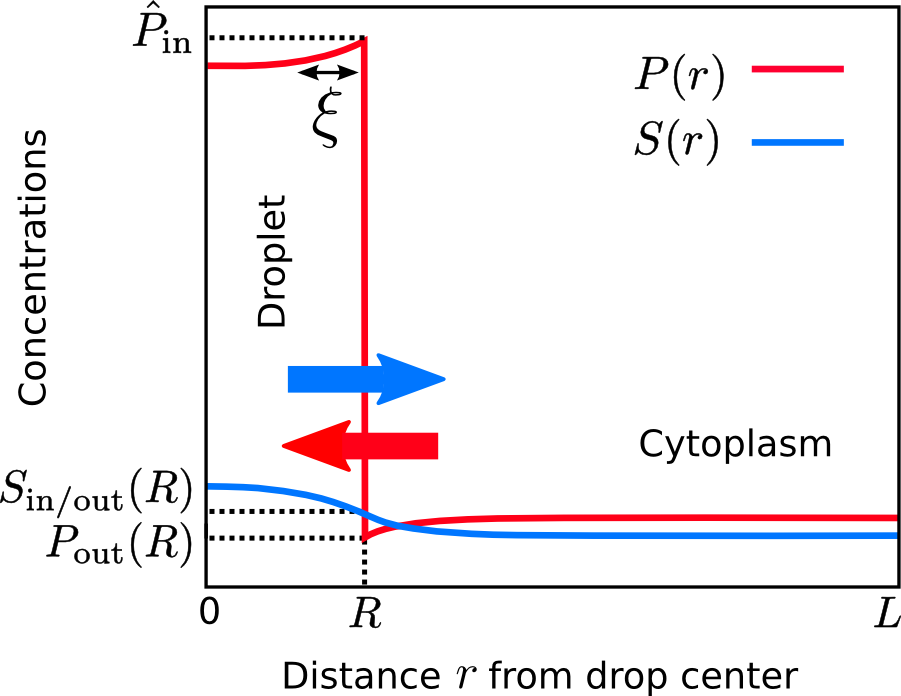}
	\caption{
	\textit{Chemical reactions induce concentration gradients inside and outside a  phase-separated droplet.} The reaction-diffusion equations (\eqs \eqref{eq:main}) solved with the local thermal equilibrium conditions at the interface (\eqs \eqref{eq:interfacePin}-\eqref{eq:NJP:S}) predict concentration gradients of length scale $\xi$ (double arrow, \eq \eqref{eq:NJP:xi}), and subsequent influx of $P$ (red arrow) and out-flux of $S$ (blue arrow) at the droplet's interface. }
	  \label{fig:NJP:profile}
\end{figure}

Setting the left hand sides of \eqs (\ref{eq:main}) to zero, the generic solutions  for $P_{\rm in/out}$, assuming spherical symmetry, are of the form \cite{wurtz_prl18}
\beqn
\label{eq:NJP:profile}
P_{\rm in/out} (r) = U^{(0)}_{\rm in/out}+\frac{R}{r}\left(U^{(1)}_{\rm in/out} \ee^{r/\xi} +U^{(-1)}_{\rm in/out}\ee^{-r/\xi}\right)  
\eeqn
where 
\beq
\label{eq:NJP:xi}
\xi  \equiv \sqrt{\frac{D}{k_i + h_i}} \ ,
\eeq
which corresponds to the length scale of the concentration gradients, and $U^{(n)}_{\rm in/out}$ are independent of $r$ but model-parameter dependent. Specifically, $U^{(n)}_{\rm in/out}$ depend on $k_i,h_i,{\hat P}_{\rm in},{\hat P}_{\rm out},l_c,L$ and $R$ as well as the overall combined concentration $\phi$ of both phase-separating ($P$) and soluble ($S$) forms of the molecules. 
Note that Eqs (\ref{eq:interfacePin})--(\ref{eq:NJP:boundary2}) together with the total mass conservation seem to suggest that there are 8 conditions to be satisfied, however, by summing the two equations in \eq (\ref{eq:main}), we see that the total protein concentration profile ($P(r)+S(r)$) in both the dilute and condensed phases have to be flat. As a result, two of the conditions in \eq (\ref{eq:NJP:boundary2}) are redundant. Therefore, we have only six conditions to be satisfied, hence the six parameters in \eq (\ref{eq:NJP:profile}). Furthermore, the concentration profiles of $S$ can be written as $S_{\rm in}(r) = \hatPin + S_{\rm in}(R_-) - P_{\rm in}(r)$ and $S_{\rm out}(r) = P_{\rm out}(R_+) + S_{\rm out}(R_+) - P_{\rm out}(r)$.


At the steady-state, $U^{(n)}_{\rm in/out}$ are determined by the boundary conditions and the molecule number conservation.
\eq \eqref{eq:NJP:profile} indicates that chemical reactions affect the concentration profiles and introduce a new length scale ($\xi$) into the problem (\fig \ref{fig:NJP:profile}).

At thermal equilibrium ($k,h=0$), a finite, phase-separating system in the nucleation and growth regime (which is the regime relevant to our biological context) can only be in two steady-states: either the system is well-mixed (i.e., no granules) or 
a single granule enriched in $P$
co-exists with the surrounding cytoplasm that is dilute in $P$ \cite{bray_advphys02}.
Furthermore, due to surface tension, there exists a critical radius $R_c$ below which droplets are no longer thermodynamically stable (\eq \eqref{eq:NJP:app:RcEq} in \ref{sec:NJP:appendix}). The critical radius can be estimated as a trade off between the surface energy ($\propto R^2$) that penalises having two phases and the bulk free energy in the droplet ($\propto R^3$) that promotes droplet formation. As a result, in the early stage of phase separation when the mixture is homogeneous,  droplets larger than $R_c$ need to be nucleated.
A quantitative analysis of nucleation is beyond the scope of this paper and we will assume that droplets are spontaneously nucleated when the system is inside the equilibrium phase boundary (Fig. 1(b)), potentially due to heterogeneous nucleation involving proteins or RNA as seeding platforms \cite{zhang_cell15, altmeyer_nature15}.
 Once multiple droplets are nucleated, droplets first grow by incorporation of phase-separating molecules $P$ that diffuse from the cytoplasm to SG. This process, known as diffusion-limited growth \cite{bray_advphys02}, stops when the cytoplasmic concentration of $P$ is depleted and reaches the saturation concentration $\hatPo$. The resulting multi-droplet system is always unstable and coarsen by Ostwald ripening \cite{lifshitz_jpcs61} -- the mechanism by which large droplets grow while small droplets dissolve -- and/or coalescence of droplets upon encountering  via diffusion \cite{siggia_pra79}.
Since  the diffusion of protein complexes in the cytoplasm is strongly suppressed \cite{weiss_biophysj04}, we will ignore droplet diffusion and coalescence completely here and focus on Ostwald ripening. The phenomenon of Ostwald ripening is caused by 1) the Gibbs-Thomson effect that dictates that the solute concentration ($P$) outside a large droplet is smaller than that outside a small droplet (\eq (\ref{eq:NJP:gt})); and 2) the dynamics of the solute in the dilute medium is well described by diffusion (\eq (\ref{eq:main}a) with $k=h=0$). These two effects combined lead to diffusive fluxes of solute from small droplets to large droplets and thus 
eventually a single droplet survives in a finite system \cite{lifshitz_jpcs61}. 

Surprisingly, Ostwald ripening can be suppressed when non-equilibrium chemical reactions are present  ($k,h>0$) \cite{glotzer_prl95,zwicker_pre15,wurtz_prl18}. 
The physical reason 
lies in the new diffusive fluxes set up by the chemical reactions (\fig \ref{fig:NJP:profile}). 
For instance, $P$ is highly concentrated inside the droplets, and gets converted to $S$ continuously in the interior. At the steady-state, the depletion of $P$ inside the droplet has to be balanced by the influx of $P$ from the outside medium.
Intuitively, we expect that the depletion of $P$ inside a droplet to scale like the droplet volume ($\propto R^3$), and the influx of $P$ through its droplet boundary to scale like the surface area ($\propto R^2$).
Taken together, we can see that as $R$ increases, the depletion rate of a droplet will eventually surpass the influx of molecules from the medium and thus the growth will stop at a steady radius. As a result, multiple droplets  of around the same size can co-exists in the system. We note that the above argument, although qualitatively correct, ignore pre-factors that also depend on the droplet radius $R$, which means that the quantitative calculation is more involved. Indeed, the influx of $P$ actually scales like $R$ instead of $R^2$ \cite{wurtz_prl18}. 

\subsection{Theoretical results summary}
\label{sec:NJP:theory}

We now briefly summarize results from Ref. \cite{wurtz_prl18} for the stability of our multi-droplet system at steady-state. The overall concentration of $P$ and $S$ in the whole system, denoted by $\bar P$ and $\bar S$ respectively, are controlled solely by the chemical reaction rates:
\beqn
\label{eq:NJP:barP}
\bar P= \frac{h_i(\alpha) }{k_i(\alpha)+h_i(\alpha)}\phi \sep \bar S= \frac{k_i(\alpha) }{k_i(\alpha)+h_i(\alpha)}\phi \ ,
\eeqn
where we recall that $\phi \equiv \bar{P} + \bar{S}$ is the overall concentration of SG constituent molecules, whether phase-separating or soluble. In particular, $\bar{P}$ depends on the chemical reaction rate constants $k_i,~h_i$, the ATP concentration $\alpha$ as well as the model under consideration (A, B or C, \eqs \eqref{eq:NJP:modelA}--\eqref{eq:NJP:modelC}). 
The qualitative behavior of the system can be categorized into regimes based on the magnitude of the droplet radius $R$ relative to the gradient length scale $\xi$ (\eq \eqref{eq:NJP:xi}).

\begin{figure}
	\centering
	\includegraphics[scale=1.2]{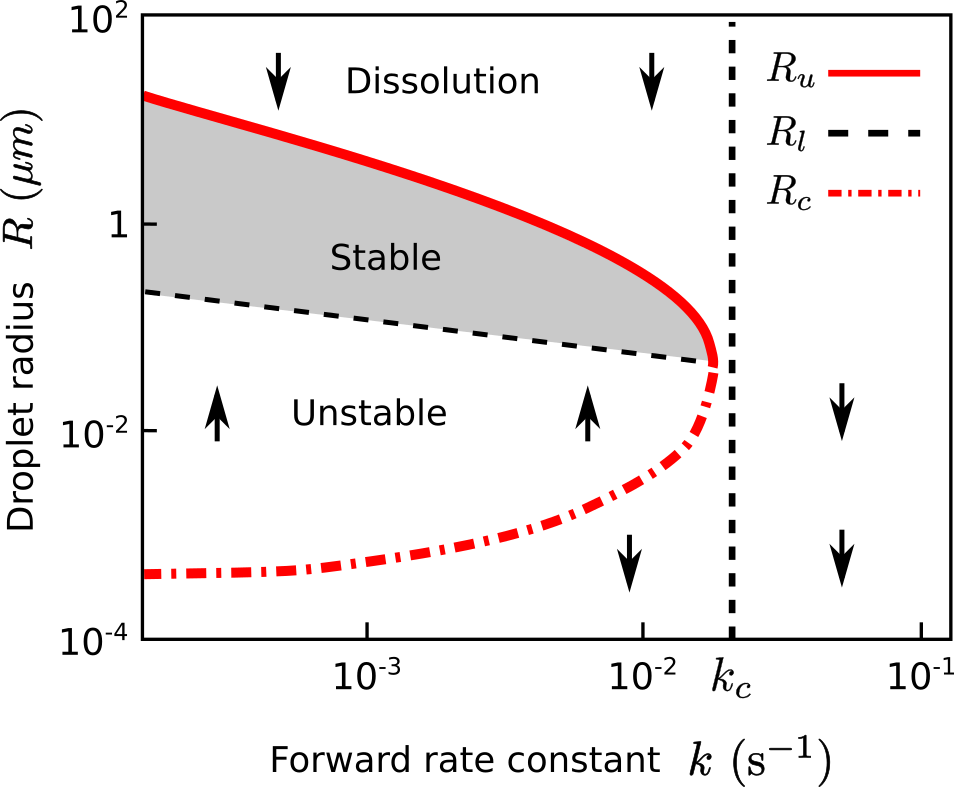}
	\caption{
		{\it Small droplet regime: chemical reactions arrest Ostwald ripening and droplet growth.} The stability of a multi-droplet system in the small droplet regime ($R\ll \xi$, Sec. \ref{sec:NJP:theory}) is shown for varying droplet radius $R$ and forward rate constant $k$. The backward rate constant $h$ is fixed and the droplet number density is variable. Droplets whose radius is smaller than $R_c$ (red irregular dashed curve, \eq \eqref{eq:NJP:theory:RcRlRu}) dissolve. A mono-disperse system is unstable against Ostwald ripening and coarsen if $R_c<R<R_l$ (black dashed line). For $R_l \leq R \leq R_u$ (red continuous curve), a mono-disperse system is stable against Ostwald coarsening (grey region). Droplets larger than $R_u$ shrink. There exists a critical rate constant $k_c$ beyond which all droplets dissolve (vertical dashed line, \eq \eqref{eq:NJP:kc}).
		Parameters: $\phi=0.2 ~\rm \mu M$, $\hatPo=0.04 ~\rm \mu M$, $\hatPin=40 ~\rm \mu M$, $l_c=1~\rm nm$, $D=1 ~{\rm \mu m^2 s^{-1}},~ h =5\times 10^{-3} ~\rm s^{-1}$. These parameters are meant to be generic in order to elucidate the system's behaviour.
	}  \label{fig:NJP:newFig}
\end{figure}

In the small droplet regime ($R \ll \xi$), the stability diagram of a monodisperse multi-droplet system, obtained from \cite{wurtz_prl18}, is shown in \fig \ref{fig:NJP:newFig} for varying droplet radius $R$, forward rate constant $k_i$, and at fixed backward rate constant $h_i$. The  droplet number density is variable in this figure, and  depends on the nucleation process and potentially the coarsening kinetics. Droplet stability depends on the droplet radius $R$ relative to the critical radii $R_c<R_l<R_u$ \cite{wurtz_prl18}:
\begin{equation}
\label{eq:NJP:theory:RcRlRu}
R_c = \frac{\hatPo l_c}{\bar P-\hatPo  } \sep R_l=\left( \frac{3 D l_c \hatPo}{2 k_i(\alpha) \hatPin} \right)^{1/3} \sep R_u=\sqrt{ \frac{3 D \left( \bar P - \hatPo  \right) }{k_i(\alpha) \hatPin}    } \ ,
\end{equation}
Droplets whose radius $R$ is smaller than the nucleus radius $R_c$ (red irregular dashed line in \fig \ref{fig:NJP:newFig}) are thermodynamically unstable and dissolve. For $R_c<R<R_l$ (black dashed line) the system is unstable against Ostwald ripening and the average droplet radius increases with time (upward arrows). 
Hence, for $R<R_l$, the system behave qualitatively like an equilibrium system (without chemical reactions). On the contrary for $R_l\leq R \leq R_u$ (continuous red line, grey region), a mono-disperse system is stable against Ostwald ripening, and for $R>R_u$ droplets shrink (downward arrows). The forward rate constant $k_i$ is bounded by the critical values $k^* < k_c$:
\beqn
\label{eq:NJP:kc}
\label{eq:NJP:ku}
k^* = \frac{2(\phi-\hatPo)}{\hatPin} h_i(\alpha)  \sep k_c = \frac{\phi - \hatPo}{\hatPo} h_i(\alpha) \ .
\eeqn
If $k_i>k_c$ all droplets dissolve due to the high conversion rate of phase-separating states $P$ into soluble states $S$ (\eq \eqref{eq:NJP:reactions}). If $k_i \lesssim k^*$ droplet radii $R$ may be comparable or larger than $\xi$. Therefore the system enters the large droplet regime where the relations in \eq \eqref{eq:NJP:theory:RcRlRu} do not apply and Ostwald ripening may not be arrested. The interested reader is referred to \cite{wurtz_prl18} for a detailed analysis of this regime.

We will now employ the formalism discussed to study the behaviour of the three minimal models introduced in Sec.~\ref{sec:NJP:models}.

\section{Model selection}
\label{sec:NJP:modelElimination}
\subsection{Model B}
\label{sec:NJP:modelB}
We will start with model B. In this model, ATP drives the $S$ to $P$ conversion (i.e., $k_B=K_B$ and $h_B=\alpha H_B$, \eq \eqref{eq:NJP:modelB}). 
 As such, a reduction in ATP will naturally suppress this conversion and thus lead to a decrease in $P$ and restrain phase separation. Therefore, depleting ATP cannot promote SG formation. As a result, we can eliminate this model since it contradicts our first biological constraint (Sec.~\ref{sec:NJP:experiments}).

\begin{figure}
	\centering
	\includegraphics[scale=0.80]{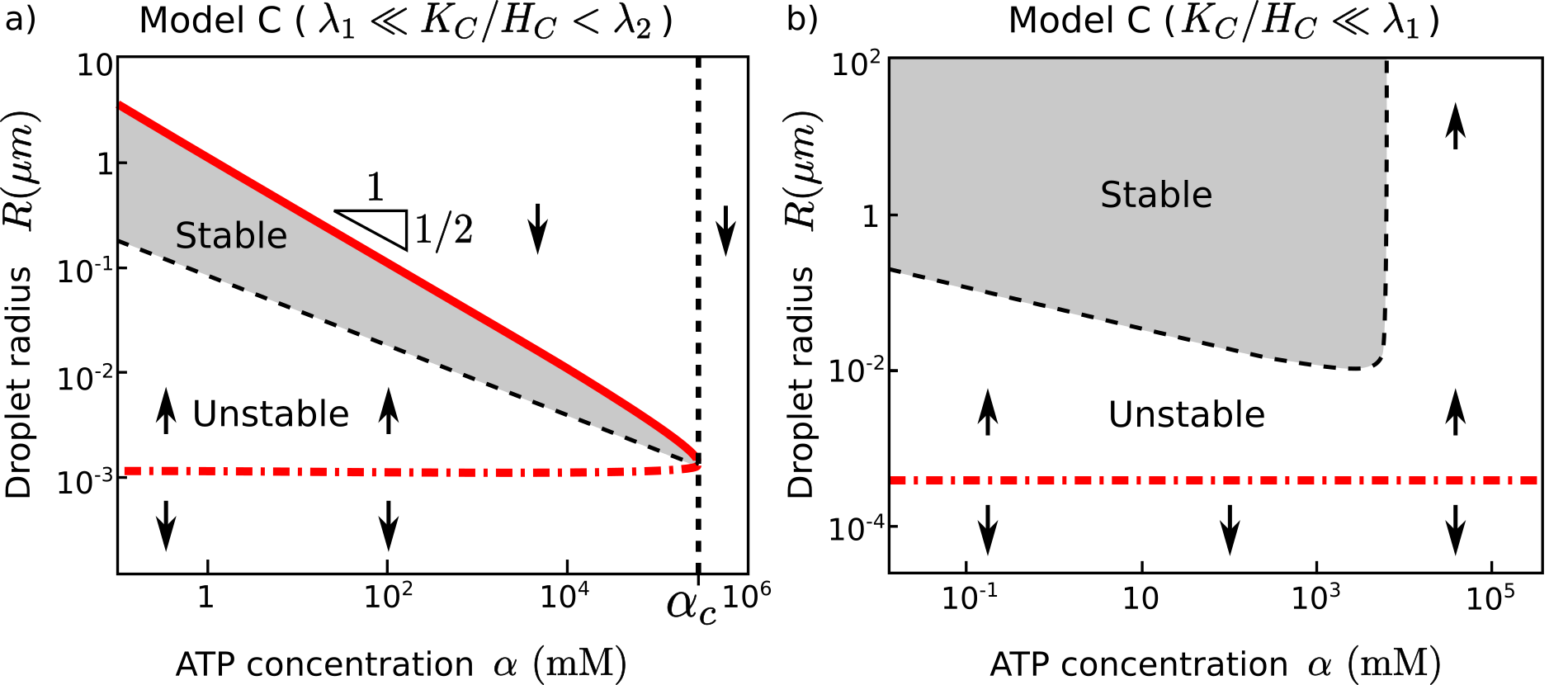}
	\caption{
	\textit{Stability diagram of model C.}
	Two regimes can be distinguished depending on the magnitude of $K_C/H_C$ with respect to the parameter $\lambda_1,~\lambda_2$ (\eq \ref{eq:NJP:lambda}). 		
	(a)   $\lambda_1 \ll K_C/H_C < \lambda_2$: droplets can exist below a critical ATP concentration $\alpha_c$ (vertical dashed line). Droplets of radius smaller than the nucleus radius $R_c$ (discontinuous red curve), or larger than the maximal radius $R_u$ (continuous red curve) are unstable and dissolve (downward arrows). Droplets larger than $R_c$ but smaller than a critical radius (black slanted  dashed line) are unstable and coarsen via Ostwald ripening, leading to an increase of the average droplet radius (upward arrows). Above the critical radius and  below $R_u$ droplets are stable (grey region). All droplet dissolve for $\alpha>\alpha_c$.
	(b) $K_C/H_C \ll \lambda_1$: $\alpha$ controls the stability of the droplets but not their formation and dissolution.
	Parameters: $\phi=0.2 ~\rm \mu M$, $\hatPo=0.04 ~\rm \mu M$, $\hatPin=40 ~\rm \mu M$, $l_c=1~\rm nm$, $D=1 ~\rm \mu m^2 s^{-1}$. (a): $K_C=5\times 10^{-3} ~\rm mM^{-1} s^{-1}$, $H_C =5\times 10^{-3} ~\rm mM^{-1} s^{-1}$. (b): $K_C=5\times 10^{-3} ~\rm mM^{-1} s^{-1}$, $H_C =10~\rm mM^{-1} s^{-1}$. These parameters are meant to be generic in order to elucidate the system's behaviour.
	}  \label{fig:NJP:modelC}
\end{figure}

\subsection{Model C}
\label{sec:NJP:modelC}
 In model C, ATP drives both conversions (i.e., $k_C=\alpha K_C$ and $h_C=\alpha H_C$, \eq \eqref{eq:NJP:modelC}). As a result, the overall concentrations $\bar{P}$ and $\bar{S}$ are independent of $\alpha$ (\eq (\ref{eq:NJP:barP})).  
 However $\alpha$ controls the multi-droplet stability as we shall see.
 We can distinguish qualitatively distinct regimes for model C depending on the relative magnitude of $K_C/H_C$ and the parameters
\beqn
\label{eq:NJP:lambda}
\lambda_1 \equiv \frac{\phi-\hatPo}{\hatPin} \sep
\lambda_2 \equiv \frac{\phi-\hatPo}{\hatPo} \ .
\eeqn

\subsubsection{$K_C/H_C > \lambda_2$ regime.}

Using \eqs \eqref{eq:NJP:modelC}, \eqref{eq:NJP:kc} and \eqref{eq:NJP:lambda}, it results that $k_C$ is greater than the critical rate $k_c$ in this regime. Therefore the system is outside the phase separating region, no droplets can form and we can discard this regime.

\subsubsection{$\lambda_1 \ll K_C/H_C < \lambda_2$ regime (\fig \ref{fig:NJP:modelC}(a)).}
\label{sec:NJP:modelCReal}

In this regime, using \eqs \eqref{eq:NJP:modelC}, \eqref{eq:NJP:kc} and \eqref{eq:NJP:lambda} leads to $k^*\ll k_C < k_c$, so droplets can form in the small droplet regime.  The multi-droplet stability is described by \eq \eq \eqref{eq:NJP:modelC} and \eqref{eq:NJP:theory:RcRlRu}. As shown in \fig \ref{fig:NJP:modelC}(a), droplets tend to shrink as the ATP concentration $\alpha$ increases,
and droplets dissolve below the nucleus radius $R_c$ (\eq \eqref{eq:NJP:theory:RcRlRu}, red irregular dashed line). Systems with droplets larger than $R_c$ are unstable against Ostwald ripening leading to an increase of the average droplet size (upward arrows). Above the critical radius $R_l$ (black slanted dashed line) there is a region where monodisperse systems are stable (grey area) and this region is bounded by the maximal radius $R_u$ (red continuous line) such that droplets larger than $R_u$ shrink (downward arrows). Specifically, $R_u$ has the following form (\eqs \eqref{eq:NJP:modelC}, \eqref{eq:NJP:barP} and \eqref{eq:NJP:theory:RcRlRu}):
\beqn
R_u=\sqrt{\frac{3 D }{\alpha K_C \hatPin}\left( \frac{H_C\phi }{K_C+H_C} - \hatPo  \right) } \ ,
\eeqn
and we therefore find the scaling form:
\beqn
\label{eq:NJP:scalingC}
R_u \propto \alpha^{-1/2} \ .
\eeqn

Namely, a fall of $\alpha$ increases the size of stable SG, thus satisfying our first constraint discussed in Sect.~\ref{sec:NJP:experiments}. However, the scaling of the maximal droplet radius $R_u$ with the ATP concentration $\alpha$ is sub-linear. For instance, to decrease the maximal SG radius $R_u$ by two-fold, a four-fold  increase in  the ATP concentration  $\alpha$ is required. Therefore, depletion of ATP according to the scaling relation in \eq (\ref{eq:NJP:scalingC}) alone cannot account for the switch-like behaviour, which is our second biological constraint. 

However, we cannot yet rule out this model because of another intriguing feature of this type of non-equilibrium phase-separating systems. 
When $\alpha$ is greater than a critical value $\alpha_c$, even though the overall concentrations $\bar{P}$ and $\bar{S}$ remain constant, one can still eliminate droplets completely by quenching the stable radii below the nucleus radius $R_c$ ($R_u<R_c$) 
(\eq \eqref{eq:NJP:theory:RcRlRu}).
An estimate of an upper bound of $R_c$ can be given by the smallest granule observed, which we take to be of the order 100 nm \cite{anderson_currbiol09}.
As we demonstrate in \ref{sec:NJP:app:C}, the scaling law (\eq \eqref{eq:NJP:scalingC}) remains valid until $R_u \simeq R_c$ so we can use it to estimate the maximal SG size that would form upon varying $\alpha$ by a factor of two in the vicinity of $\alpha_c$. As a conservative estimate, if we assume that the tip of the phase boundary where $\alpha=\alpha_c$ (\fig \ref{fig:NJP:modelC}(a)) corresponds to the ATP concentration in normal conditions, and that the corresponding droplet size is $R_c$, then a reduction of 50\%  of $\alpha$ can only lead to a maximal SG radius of around $100 \times 2^{1/2}\simeq 140 ~\rm nm$  according to the upper bound law $R_u$. This radius is too small compared to experimental observations (Sec.~\ref{sec:NJP:experiments}) and we thus rule out model C in this regime.

\subsubsection{$K_C/H_C \ll \lambda_1$ regime (\fig \ref{fig:NJP:modelC}(b)).}

In this regime, we find from \eqs \eqref{eq:NJP:modelC}, \eqref{eq:NJP:ku} and \eqref{eq:NJP:lambda} that $K_C \ll k^*$, therefore droplets may be larger than the gradient length scale $\xi$ (Sec. \ref{sec:NJP:theory}). 
As shown in \fig \ref{fig:NJP:modelC}(b), the ATP concentration $\alpha$ also controls the droplet stability but droplets have unbounded  radii. In addition, irrespective of $\alpha$, an existing droplet emulsion  either remains stable (grey region), or droplets coarsen via Ostwald ripening (upward arrows), i.e., droplets do not dissolve away. 
Since one cannot control droplet assembly and dissolution based on the magnitude of $\alpha$ we can eliminate model C in this regime as well.

In summary, we have shown that model C does not provide the switch-like response compatible with our second biological constraint  (Sec.~\ref{sec:NJP:experiments}). We can therefore rule out this particular model.

\subsection{Model A}
\label{sec:NJP:modelA}

\begin{figure}
	\centering
	\includegraphics[scale=0.80]{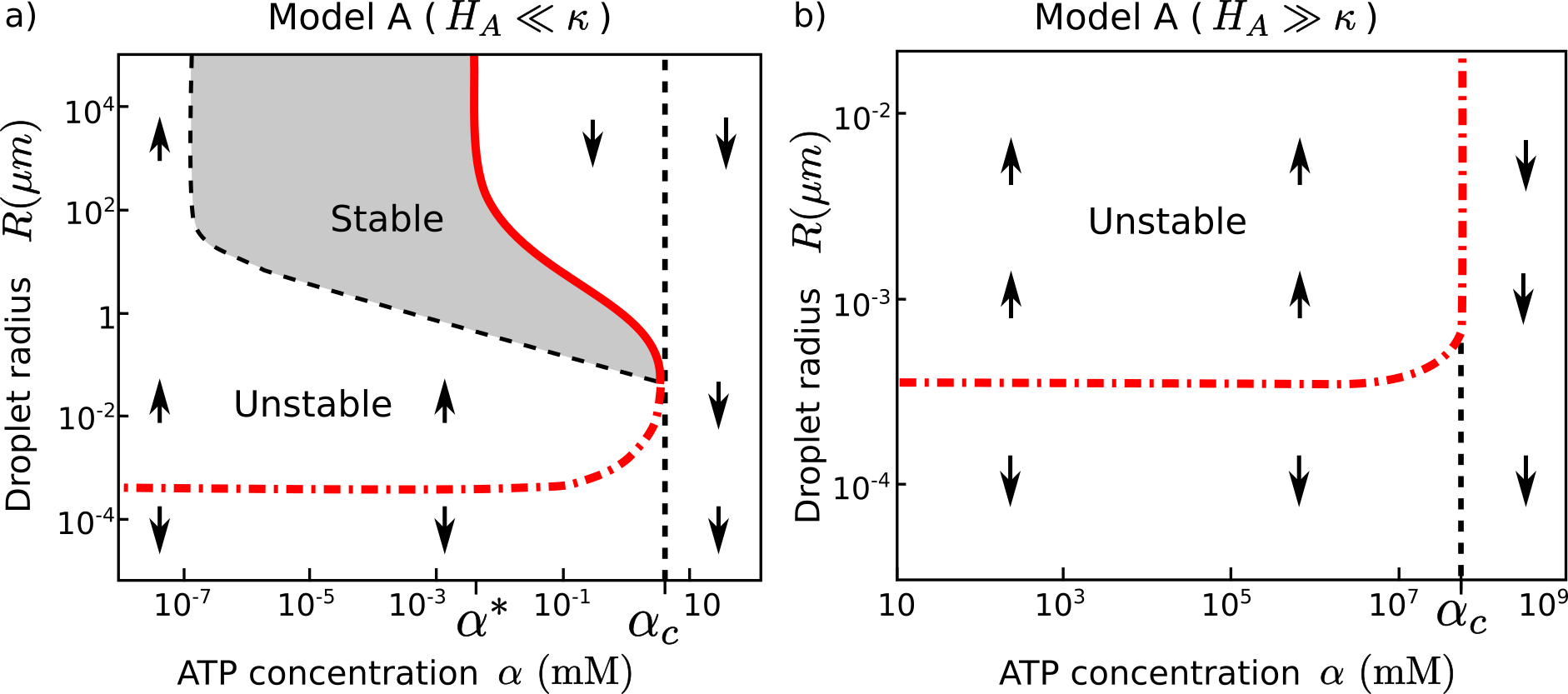}
	\caption{
	\textit{Stability diagram of model A.} 
	Two regimes can be distinguished depending on the magnitude of  $H_A$ with respect to the parameter $\kappa$ (\eq \ref{eq:NJP:kappa}).
	(a) $H_A \ll \kappa$: droplets can exist below a critical ATP concentration $\alpha_c$ (black vertical dashed line). 
		Droplets of radius smaller than the nucleus radius $R_c$ (discontinuous red curve), or larger than the maximal radius $R_u$ (continuous red curve) are unstable and dissolve (downward arrows). Droplets larger than $R_c$ but smaller than a critical radius (black dashed curve) are unstable and coarsen via Ostwald ripening, leading to an increase of the average droplet radius (upward arrows). There exists another critical ATP concentration $\alpha^*$ 
			below which droplets radius are not maximally bounded.  For $\alpha \gg \alpha^*$ (\eq \eqref{eq:NJP:ku}), $R_c,~R_l$ and $R_u$ are well approximated by \eq \eqref{eq:NJP:theory:RcRlRu}.
	(b)  $H_A \gg \kappa$: droplets can exist only  below a critical ATP concentration $\alpha_c$ 
	 and are always unstable and coarsen \textit{via} Ostwald ripening (upward arrows).
	Parameters: $\phi=0.2 ~\rm \mu M$, $\hatPo=0.04 ~\rm \mu M$, $\hatPin=40 ~\rm \mu M$, $l_c=1~\rm nm$, $D=1 ~\rm \mu m^2 s^{-1}$, $K_A=5\times 10^{-3} ~\rm mM^{-1} s^{-1}$. (a): $H_A =5\times 10^{-3} ~\rm s^{-1}$. (b): $H_A =5\times 10^{7} ~\rm s^{-1}$. These parameters are meant to be generic in order to elucidate the system's behaviour.
	\label{fig:NJP:modelA}		
	}  
\end{figure}

In model A only the forward reaction rate $k$ is amplified by an increase of the ATP concentration $\alpha$ and the backward rate $h$ is constant (i.e., $k_A=\alpha K_A$ and $h_A= H_A$, \eq \eqref{eq:NJP:modelA}). 
We can again distinguish two qualitatively distinct regimes, depending on the magnitude of $H_A$ relative to the parameter
\beqn
\label{eq:NJP:kappa}
\kappa \equiv \frac{D}{l_c^2} \left(\frac{\phi-\hatPo}{\hatPo}\right)^2
\ .
\eeqn


\subsubsection{$H_A\gg \kappa$ regime (\fig \ref{fig:NJP:modelA}(b)).}
In this regime, it can be shown that the nucleus radius $R_c$ (\eq \eqref{eq:NJP:theory:RcRlRu}) is necessarily larger than the gradient length scale $\xi$ (\eq \eqref{eq:NJP:xi}). Therefore,  since stable droplets are always larger than the nucleus, all droplets are also larger than $\xi$ and the system is in the large droplet regime (Sec. \ref{sec:NJP:theory}). 
As shown in \fig \ref{fig:NJP:modelA}(b), Although there exists a critical ATP concentration $\alpha_c$ beyond which all droplets dissolve (vertical dashed line), a multi-droplet system is always unstable and coarsen via Ostwald ripening (upward arrows). This is not a desirable feature for the formation of stable cytoplasmic organelles and we therefore discard model A in this regime.

\subsubsection{$H_A \ll \kappa $ regime (\fig \ref{fig:NJP:modelA}(a))}.
\label{sec:NJP:modelAa}

As can be seen on \fig \ref{fig:NJP:modelA}(a), two qualitatively distinct regions can be distinguished based on the ATP concentration $\alpha$. When $\alpha$ is much smaller than the critical value $\alpha^* \equiv \lambda_1 H_A/K_A$ (\eq \eqref{eq:NJP:lambda}), it can be seen from \eqs \eqref{eq:NJP:modelA}, \eqref{eq:NJP:ku} and \eqref{eq:NJP:lambda} that $k_A \ll k^*$.  Hence large droplets may exist in the system (Sec. \ref{sec:NJP:theory}) and no upper-bound radius exists. 

 If on the contrary $\alpha\gg \alpha^*$, then $k_A\gg k^*$ so the system is in the small droplet regime (Sec. \ref{sec:NJP:theory}). Therefore, the nucleus radius $R_c$ (red irregular dashed curve), the instability-stability transition radius $R_l$ (black dashed curve) and the maximal droplet radius $R_u$ (continuous red curve) are well approximated in this region by \eq \eqref{eq:NJP:theory:RcRlRu}.
In particular, Using  \eqs \eqref{eq:NJP:modelA}, \eqref{eq:NJP:barP} and \eqref{eq:NJP:theory:RcRlRu}, the expression of the maximal droplet radius $R_u$ is given by:
\beqn
\label{eq:NJP:scalingA:bis}
R_u =  \sqrt{\frac{1}{\alpha} \left(\frac{H_A \phi}{\alpha K_A+H_A}- \hatPo\right)}
\ ,
\eeqn
Similar to Model C with $\lambda_1 \ll K_C/H_C < \lambda_2$ (Sec. \ref{sec:NJP:modelCReal}), the maximal droplet radius $R_u$ decreases with the ATP concentration $\alpha$. A major difference however, is that the shrinkage is more pronounced due to the additional $\alpha$ in the denominator of $\bar P$ (\eq \eqref{eq:NJP:barP}). This is due to the fact that increasing $\alpha$ accelerates the conversion $P \rightarrow^k S$, but not the reverse reaction. This leads to a reduction of the overall concentration $\bar P$ of phase separating-material in the system (\eq \eqref{eq:NJP:barP}). 

When $\alpha$ is greater than the critical value $\alpha_c$ (vertical dashed line), $k_A>k_c$ (\eq \eqref{eq:NJP:kc}) and all droplets dissolve ($R_u=0$). This is due to the system being driven outside the phase-separating region by the chemical reactions ($\bar P \leq \hatPo$, \fig \ref{fig:NJP:fig1}(b), `$\square$' symbol). 
We find the expression of $\alpha_c$ by solving $k_A(\alpha_c)=k_c$:
\beqn
\label{eq:NJP:modelA:alpha_c}
\alpha_c=\left(\frac{\phi}{\hatPo}-1\right) \frac{H_A}{K_A} \ .
\eeqn

As a result, in model A droplet dissolution can be achieved by depleting $P$ to the extent that the system crosses the equilibrium phase boundary. This suggests a stronger response than in model C which we quantify in \ref{sec:NJP:app:A}. In particular we show that the ratio $R_u/R_c$ for $\alpha \lesssim \alpha_c$ 
does not have to be of order $1$ as in model C but is a function of the system parameters $\hatPo,\hatPin,\phi,D,l_c$ and $H_A$. Therefore in the $H_A \ll \kappa$ regime of model A,  droplets can be formed in a switch-like manner by a two-fold decrease of $\alpha$, satisfying both our  biological constraints (Sec.~\ref{sec:NJP:experiments}).

In summary, we  conclude that among the three minimal models introduced, model A in the $ H_A \ll \kappa$ regime is the  best suited to describe the physics of ATP-triggered SG formation.

\section{Model comparison to experimental observations}

 We are now left with a unique model, Model A with the backward rate constant $H_A$ such that $H_A\ll \kappa$ (\eq \eqref{eq:NJP:kappa}, Sec. \ref{sec:NJP:modelAa}). Using physiologically relevant parameters determined in \ref{sec:NJP:param} we now compare our model predictions to experimental observations on SG.

\subsection{SG steady-state}
\label{sec:NJP:steadystate}

\begin{figure}
	\centering
	\includegraphics[scale=1]{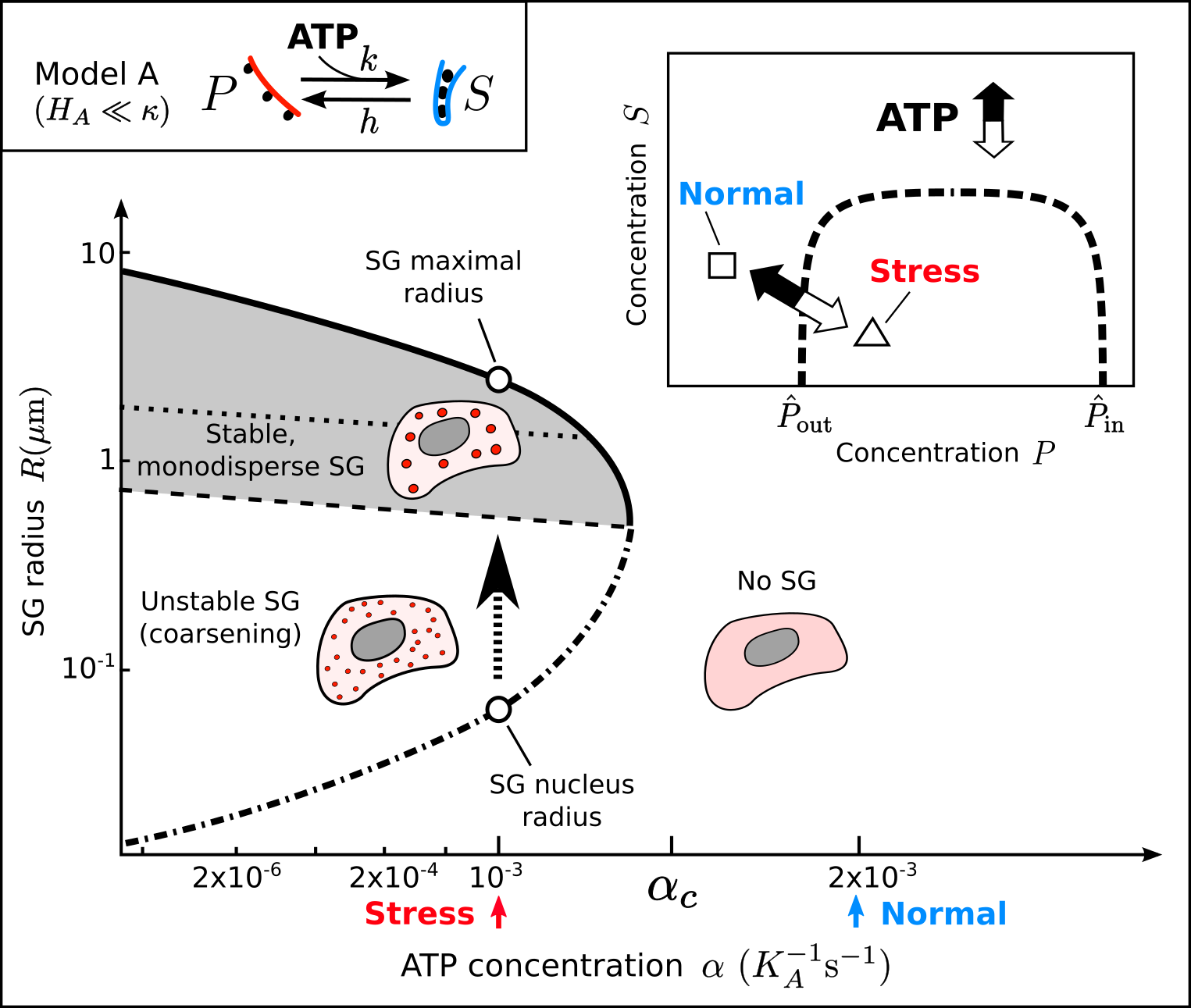}
	\caption{
		Among the three minimal models proposed, only model A in the $H_A \gg \kappa$ (\eq \eqref{eq:NJP:kappa}) regime can describe SG formation and dissolution upon two-fold variations of the ATP concentration $\alpha$. In this model phase-separating states $P$ are converted into soluble states $S$ at a rate proportional to $\alpha$. During normal conditions $\alpha$ is high (blue arrow) and there are few molecules $P$ so that the system does not phase separate (`$\square$' symbol in insert). When $\alpha$ reduces two-fold during environmental stresses (red arrow) the concentration of $P$ increases, making the system cross the phase boundary and SG assemble by phase separation (`$\triangle$' symbol).  SG nucleate from small droplets of radius $R_c$ of about $50 ~\rm nm$ then grow  and/or coarsen (black upward arrow) until they reach a stable radius $R_u$ (\eq \eqref{eq:NJP:scalingA:bis}) between $0.5$ and $2.2 \rm \mu m$ (grey region). The radius $R_s$ above which 
			the spherical shape is unstable (\eq \eqref{eq:NJP:Rdiv}) is shown by the dotted line. 
		Parameters are from \ref{sec:NJP:param} and $\hatPo$ and $K_A$ can be chosen arbitrarily.
		\label{fig:NJP:lastFig}
	}  
\end{figure}

We show in \fig \ref{fig:NJP:lastFig} the SG stability diagram for varying ATP concentration $\alpha$ and droplet radius $R$. In normal conditions $\alpha$ is at its basal value (blue arrow) leading to a low overall concentration $\bar P$ of phase-separating material. As a result, the system is outside the phase-separating region ($\bar P < \hatPo$, `$\square$' in the insert figure) and no droplets can exist. During stress conditions, ATP decreases two-fold (red arrow) leading to an increase of $\bar P$, thus taking the system inside the phase-separating region ($\bar P>\hatPo$, `$\triangle$'). SG  form via the nucleation of small droplets of radius  $50 ~\rm nm$. Nuclei then grow and/or coarsen (upward black arrow) leading to stable SG with much larger radii, between $0.5 \rm \mu m$ and $2.3 \rm \mu m$ (grey region), depending on the number of nucleated droplets. The relationship between the number of nucleation events and the radius of stable SG will be investigated in Sec.~\ref{sec:NJP:dynamics}. We note that as ATP increases and crosses the dissolution point $\alpha_c$ (\eq \eqref{eq:NJP:modelA:alpha_c}), the radius of the smallest stable droplet deceases from $0.5$ to $0 \mu m$. Therefore our model predicts that during normal conditions SG do not only shrink below detectable levels but fully dissolve.

We thus find that the two biological constraints that SG must form when the ATP concentration decreases by two-fold, and in a switch-like manner (Sec.~\ref{sec:NJP:experiments}), are both satisfied. Furthermore, the quantitative predictions for the stable SG radii are consistent with experimental observations. Therefore, model A reproduces salient experimental observations of SG formation and dissolution based on the ATP level.

\subsection{Stability of SG shape}
\label{sec:NJP:shape}

Chemical reactions may destabilize the spherical shape of large droplets, leading to deformation or division into smaller droplets. This phenomenon, studied in a binary mixture  \cite{zwicker_natphys16} ($P$, $S$ and no cytoplasm), is caused by the gradient of the $P$ concentration being more pronounced near bulging regions, leading to a larger influx of $P$ molecules in these regions. Hence, bulging regions grow by accumulation of material $P$, while depressed regions recede. This accentuates the shape deformation 
and can cause droplet division.

The concentration profiles in our ternary system (with cytoplasm), in the regime relevant for SG formation ($H_A \ll \kappa$, Sec. \ref{sec:NJP:modelAa}), are identical to the profiles in the binary case \cite{wurtz_prl18}. We can therefore extend the results from \cite{zwicker_natphys16} to our system and determine the radius $R_{\rm  s}$ above which the droplet spherical shape becomes unstable:
\beqn
\label{eq:NJP:Rdiv}
 R_{\rm  s} &=& \left( \frac{33 D  l_c \hatPo }{k \hatPin }\right)^\frac{1}3 \ .
\eeqn

Using the parameters determined in \ref{sec:NJP:param} we 
show $R_s$ in \fig \ref{fig:NJP:lastFig} (dotted line). 
If SG are sufficiently large and the ATP is sufficiently depleted, droplets may deform and/or divide. We note however that, to our knowledge, no experimental work reported significant SG deformation or division during the stress phase.

	

So far we have restricted our analysis to the steady-states and stability of SG and ignored the dynamics of SG formation, which we will now analyse.

\subsection{SG growth and coarsening dynamics}
\label{sec:NJP:dynamics}

\begin{figure}
	\centering
	\includegraphics[scale=0.77]{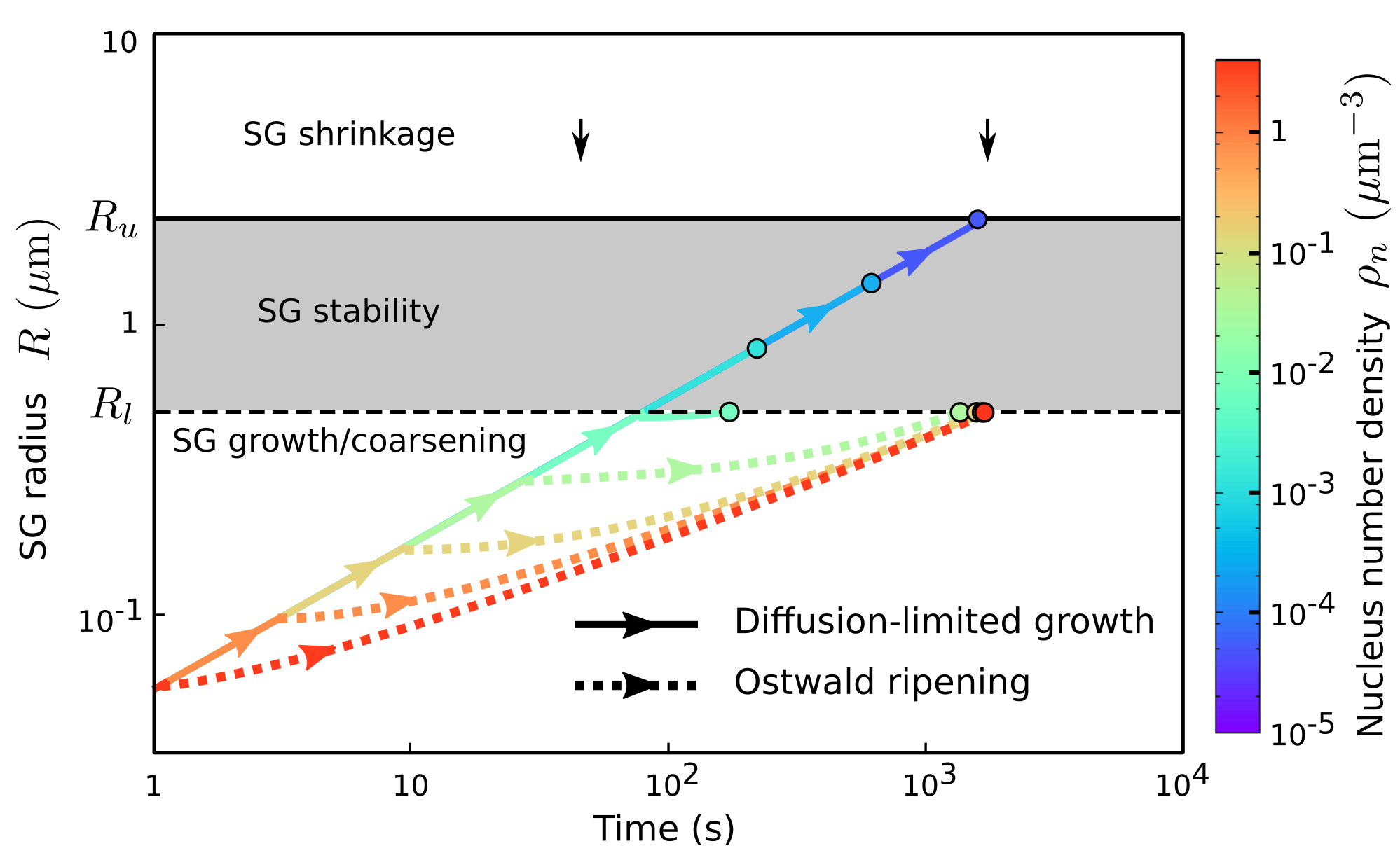}
	\caption{
		\textit{The growth and coarsening dynamics of an assembly of SG depend on the number density $\rho_n$ of nucleated droplets (\eq  \eqref{eq:NJP:dynamicsFull}).} For large $\rho_n$ SG first grow  by diffusion-limited growth (DLG) (continuous curves) then coarsen by Ostwald ripening (OR) (dotted curves). The steady-state (coloured circles) is reached when the SG radius equals $R_l$ (vertical dashed line), above which OR is arrested. For small $\rho_n$, SG grow by DLG growth only, and at steady-state the SG radius depends of $\rho_n$ and is comprised between $R_l$ and $R_u$ (grey region). Droplets larger then $R_u$ shrink (downward arrows). Parameters are from \ref{sec:NJP:param}.
	}
	\label{fig:NJP:dynamics}  
\end{figure}

Let us consider the small droplet regime, which is the relevant regime for SG formation (model A with $H_A \ll \kappa$, Sec. \ref{sec:NJP:modelAa}). As demonstrated in \ref{sec:NJP:appendix:equivalence}, the dynamics of droplet growth far from the steady-state is identical to that in an equilibrium system, i.e. without chemical reactions but with the same amount of $P$ and $S$.
The dynamics of SG formation can therefore be approximated by the dynamics of a multi-droplet system in equilibrium conditions 
(Sec.~\ref{sec:NJP:steadyStateAndStability}), 
 with the additional constraint that  Ostwald ripening (OR) and diffusion-limited growth (DLG) are in such a way that they are arrested for droplet radii larger than $R_l$ and $R_u$, respectively (\eq \eqref{eq:NJP:theory:RcRlRu}). We ignore for simplicity the nucleus radius ($R_c=0$) and the effect of droplet deformation at large radii (Sec. \ref{sec:NJP:shape}). In the equilibrium regimes (DLG and OR), at large times and far from the DLG-OR transition, the dynamics of the average droplet radius $R$ is given by 
\cite{pitaevskii_b81}:
\beqn
\label{eq:NJP:dynamics}
R_{\rm \scriptscriptstyle DLG} =(\beta_{\rm \scriptscriptstyle DLG}  t)^\frac{1}{2} \sep 
R_{\rm \scriptscriptstyle OR} =(\beta_{\rm \scriptscriptstyle OR}  t)^\frac{1}{3} \ ,
\eeqn
with $\beta_{\rm \scriptscriptstyle DLG}=2 D(\bar P-\hatPo)/\hatPin$ and $\beta_{\rm \scriptscriptstyle  OR}=D l_c \hatPo/\hatPin$. We neglect the deviation from these laws at the DLG-OR transition, and assume that the system switch from the DLG regime to the OR regime at a critical average droplet radius $R^*$ and time $t^*=(R^*)^2/\beta_{\rm \scriptscriptstyle DLG}$.
The dynamics for $t>t^*$ is therefore obtained by using the function $R_{\rm \scriptscriptstyle OR}(t)$ with the appropriate shift of the time variable $t$ so that  $R(t^*) = R^*$. We obtain the following dynamics:
\begin{equation}
\label{eq:NJP:dynamicsFull}
R(t) = \left\{
\begin{array}{ll}
 (\beta_{\rm \scriptscriptstyle DLG} t)^{\frac{1}{2}}  \quad\quad\quad\quad\quad\quad\quad\quad\quad\quad\quad\quad\quad\quad\quad\quad\quad\quad\quad\quad {\rm if~} t\leq t^*, ~R < R_u \\
 \left[ \beta_{\rm \scriptscriptstyle OR} \left( t-t^*+T \right) \right]^{\frac{1}3} 
 = \left[ \beta_{\rm \scriptscriptstyle OR} \left( t - \dfrac{(R^*)^2}{\beta_{\rm \scriptscriptstyle DLG}} +  \dfrac{(R^*)^3}{\beta_{\rm \scriptscriptstyle OR}}  \right )\right]^{\frac{1}{3}} \quad {\rm if~} t > t^* , ~R < R_l
\end{array}
\right.
\end{equation}
where $T=(R^*)^3/\beta_{\rm \scriptscriptstyle OR}$ is the solution of $R_{\rm  \scriptscriptstyle OR}(T)=R^*$, and the upper-bounds to $R$ account for the stabilizing effect of the chemical reactions.

We now estimate the critical radius $R^*$ at the DLG-OR transition. In the DLG regime droplets grow by depleting the cytoplasm from phase-separating material $P$. The droplet growth in this regime stops when the cytoplasmic concentration reaches the saturation concentration $\hatPo$ (\eq \eqref{eq:NJP:gt} and \fig \ref{fig:NJP:fig1}). Since the total number of molecules $P$ is constant, $R^*$ is given by
\beqn
R^*  = \left( \frac{3 (\bar P - \hatPo)}{4 \pi \rho_n \hatPin} \right)^{\frac{1}{3}} \ ,
\eeqn
 where we used the strong phase separation approximation $\hatPin \gg \hatPo$ and where $\rho_n$ is the density number of nucleated droplets. $\rho_n$ is also equal to the droplet number density throughout the DLG regime since we have assumed that droplet coalescence does not occur in our system (Sec. \ref{sec:NJP:steadyStateAndStability}). 

As shown in \fig \ref{fig:NJP:dynamics} for the parameters determined in \ref{sec:NJP:param}, the SG formation dynamics and the steady-state radius $R$ (coloured circles) depend on the  nucleus number density $\rho_n$. For large $\rho_n$, droplets grow first by DLG (solid coloured lines), then coarsen by OR (dotted coloured curves). The steady-state is reached when droplet radii are equal to $R_l$ (vertical dashed line), above which OR is arrested. For smaller $\rho_n$, SG radius $R$ increases by DLG until $R_l<R<R_u$ (vertical solid line) where OR is not active (grey region). Therefore the system is in a stable steady-state after the DLG. For even smaller $\rho_n$ the DLG continues until droplet radii are equal to $R_u$ (vertical continuous line), which is  the maximal radius attainable. In this case, stable SG radii are equal to $R_u$ and are independent on the number of SG in the system. The critical nucleus number density $\rho_n^*$ below which droplets grow exclusively by DLG is the solution of $R^*(\rho_n^*)=R_l$ (\eq \eqref{eq:NJP:theory:RcRlRu}):
\beqn
\label{eq:NJP:rhonstar}
\rho^*_n = \frac{\left( \bar P - \hatPo \right)K_A \alpha }{2 \pi D l_c \hatPo} \ .
\eeqn
Using the physiologically relevant parameters estimated in \ref{sec:NJP:param} we find $\rho_n^*\approx 6 \times 10^{-3} \rm \mu m ^{-3}$. 

The time needed for nuclei growing exclusively in the DLG regime ($\rho_n < \rho_n^*$) to reach the radii $R_l \approx 0.5 \rm \mu m$ and $R_u \approx 2.2 \rm \mu m$ are approximatively $2 \rm min$ and $25 \rm min$, respectively. For nuclei growing exclusively in the OR regime ($\rho_n \gg \rho_n^*$), the time needed to reach the radius $R_l \approx 0.5 \rm \mu m$ is approximatively $30 \rm min$. Therefore growth by DLG and OR are both consistent with SG formation times scales.

\section{Summary \& Discussion}
\label{sec:NJP:discussion}
Starting from experimental observations of SG formation in the cell cytoplasm, we have formulated three minimal models based on 
chemical reaction-controlled phase separation to account for the appearance of SG upon ATP depletion. Applying the formalism developed in \cite{wurtz_prl18}, 
 we  compared the models based on their qualitative features  to salient experimental observations. 

We eliminated model B because it does not predict SG growth  when ATP concentration falls. Model C was discarded because although SG grow during ATP depletion, the response is not switch-like. Finally, we found that model A, where ATP drives only the $P \to S$ conversion, can satisfy both biological constraints. However, we have ruled out the $H_A\gg \kappa$ (\eq \eqref{eq:NJP:kappa}) regime in model A because droplets cannot be stable as they always coarsen via Ostwald ripening.
Using physiological parameters (\ref{sec:NJP:param}), we elucidated a particular scenario of our model, and showed that model A reproduces salient experimental observations of SG formation and dissolution based on the ATP level. Finally we analysed the SG formation dynamics and quantified the duration of growth and coarsening,  depending on the number of nucleated droplets.

A peculiar feature of model A is that under normal condition, ATP is continuously hydrolysed to keep SG from forming.
Superficially, it may seem wasteful energetically. However, this is in fact not dissimilar to any insurance schemes that we are familiar with. For instance, we pay a car insurance premium every month so that when an accident occurs, the damage cost is covered. Similarly, in our model, ATP is used during normal condition so that, during stress, no additional ATP is required for SG formation. This perspective is particularly pertinent for SG regulation since the timing of environmental stresses can be unpredictable. Furthermore, due to physical constraints such as cell size, storage of ATP for a long period of time is difficult. It may therefore be desirable to have survival mechanisms, such as SG formation, that are spontaneous and do not require additional ATP consumption for the formation of SG. 

Indeed, there is already experimental evidence suggesting that  ATP can promote SG disassembly 
by ATP-dependent protein phosphorylation, via the activity of focal	adhesion kinase (FAK) \cite{tsai_embo08}, Casein Kinase 2 \cite{reineke_mcb17} and dual specificity kinase DYRK3 \cite{wippich_cell13}.
On the other hand, while energy depletion often accompanies stress conditions \cite{lilly_cancerres84, chang_ajp01}, SG formation \cite{hofmann_mboc12} or cause SG formation \cite{kedersha_mboc02}, ATP may also be necessary to SG assembly in some situations \cite{jain_cell16} 
	We note as well that there is substantial evidence that energy depletion-independent pathways may also exist, such as via the  phosphorylation of eukaryotic initiation factor alpha (eIF2-alpha)  \cite{mcewen_jbc05}.
Given all these evidence, the cell seems to have multiple mechanisms to ensure SG assembly and disassembly at stressful times \cite{dang_jbc06} and our work may describe a particular pathway of SG regulation. Nevertheless, the virtue of our model is that it leads to specific predictions that can be tested experimentally, which we will now enumerate.

%

%
%
%
%

\subsection{Predictions}
\label{sec:NJP:predictions}
Our model provides the following experimental predictions:
\begin{enumerate}
	\item
	Since  the $P\rightarrow S$ reaction is the one that requires the input of ATP, it is natural to relate the conversion to  the ubiquitous ATP-driven phosphorylaton reaction. In other words, our model suggests that the soluble state of the 
	SG assembling constituents
	corresponds to the phosphorylated form of these constituents. 
	\item
	We predict the existence of a concentration gradient of the phase-separating constituent $P$ 
	 outside the SG, with the gradient length scale of the form
	\beq
	\xi= \sqrt{\frac{D}{\alpha K_A+H_A}}
	\ .
	\eeq
	Using biologically relevant parameters determined in \ref{sec:NJP:param} we find $\xi \approx 15 \rm \mu m$, which is relevant in the context of the cell.

	
	\item
	If the number density of nucleated droplets is large  ($\rho_n \gg \rho_n^*$, \eq  \eqref{eq:NJP:rhonstar}), we predict a relationship between the stable SG radius $R_l$ and the ATP level $\alpha$  (\eq \eqref{eq:NJP:theory:RcRlRu} and \eqref{eq:NJP:modelA}):
	\beqn
	R_l \sim \alpha^{-1/3} \ .
	\eeqn

	\item
	Finally, if ATP is depleted beyond physiological conditions and SG are large enough (small droplet number density), we predict non-spherical SG or SG splitting due to shape instabilities (Sec. \ref{sec:NJP:shape}).


\end{enumerate}
The first prediction may be tested by screening the purified constituents of SG in 
an {\it in vitro} setting. The second, third and fourth predictions can be tested using imaging techniques with well regulated ATP concentration either {\it in vivo} or {\it in vitro}. 

\appendix

\section{Largest SG to nucleus size ratio $R_u/R_c$ close to the droplet dissolution point $\alpha\simeq\alpha_c$}
\label{sec:NJP:appendix}
In model A for $H_A\ll \kappa$ and model C for $K_C/H_C \gg \lambda$ we have seen that there exists a critical ATP concentration $\alpha_c$ beyond which no droplets can exist, and  below which droplets can nucleate from a minimal radius $R_c$ and grow until they reach their stable radius that is upper bounded by $R_u$. Therefore one can control droplet formation and dissolution via small variations of $\alpha$ in the vicinity of $\alpha_c$ (\figs \ref{fig:NJP:modelC}(a) and \ref{fig:NJP:modelA}(a)). From experimental observations we know that during stress  $\alpha$ vary by two-fold and $R_u \gg R_c$, providing a constraint on our modelling (Sec.~\ref{sec:NJP:experiments}). Here we examine this constraint by quantifying the size ratio $R_u/R_c$ in the vicinity of $\alpha_c$.

\begin{figure}
	\centering
	\includegraphics[scale=1.5]{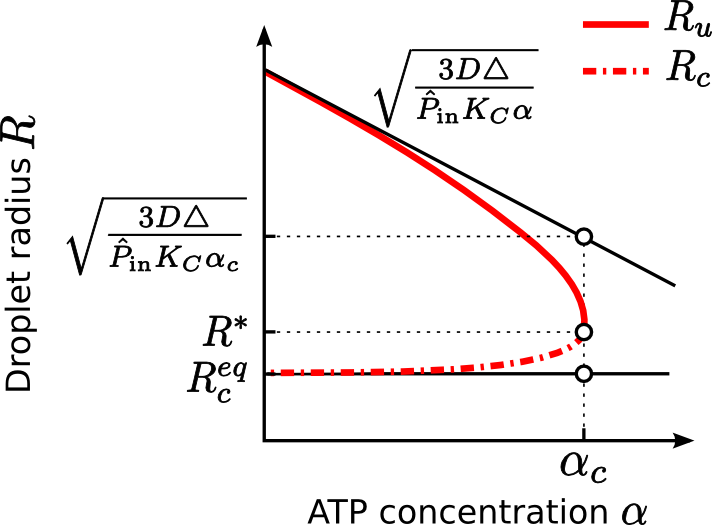}
	\caption{
		Fixed points of the droplet growth rate \eq \eqref{eq:NJP:app:J} in an infinite single-droplet system for varying ATP concentration $\alpha$, for model C in the $\lambda_1 \ll K_C/H_C < \lambda_2$ regime. The stable droplet radius $R_u$ (red continuous line) and the nucleus radius $R_c$ (red dashed line) correspond to the maximal stable radius and nucleus radius in a multi-droplet system (\fig \ref{fig:NJP:modelC} (a)).
		No droplets exist for $\alpha$ larger than a critical value $\alpha_c$. For $\alpha \ll \alpha_c$, $R_u\simeq \sqrt{3D \Delta /(\hatPin K_C \alpha )}$ (upper black line , \eq \eqref{eq:NJP:app:Ru}) and $R_c\simeq R_c^{\rm eq}$ (lower black line, \eq \eqref{eq:NJP:app:Rc}). When $\alpha \lesssim \alpha_c$, small variations of $\alpha$ lead to strong variations of $R_u$ and $R_c$. In this strong response regime the ratio $R_u/R_c$ is bounded by $\sqrt{3D \Delta /(\hatPin K_C \alpha_c)}/R_c^{\rm eq}$.
		\label{fig:NJP:appendix}
	}  
\end{figure}

We consider a single droplet of radius $R$ growing in a large system, such that the concentration far from the droplet is not affected by the presence of the droplet. Therefore the supersaturation  in the cytoplasm is constant:  $\Delta=\bar P- \hatPo$.
We seek the nucleus and maximal droplet radii $R_c$ and $R_u$ in this single-droplet system, which are identical to those in a multi-droplet system.
When droplets are much smaller than the gradient length scale $\xi$ which is true in the regimes under consideration, the net flux $J$ of molecules $P$ at the droplet's interface is composed of an influx from the medium and an out-flux due to the chemical conversion $P \to^k S$ inside the droplet that depletes $P$ \cite{wurtz_prl18}:
\beqn
\label{eq:NJP:app:J}
J= 4 \pi D R \left(\Delta - \frac{ \hat{P}_{\rm out}l_c}{R}\right)-\frac{4 \pi R^3}{3}k \hat{P}_{\rm in} \ ,
\eeqn
where the supersaturation $\Delta$ is  set by the chemical reaction rates $k,h$ (\eq \eqref{eq:NJP:barP}):
\beqn
\label{eq:NJP:triangle}
\Delta=\frac{\phi}{1+k/h}-\hat{P}_{\rm out} \ .
\eeqn
The droplet grows when $J>0$ and shrink otherwise. At equilibrium ($k=h=0$ but $k/h$ is still defined by \eqs \eqref{eq:NJP:modelA} or \eqref{eq:NJP:modelC}) there is a unique fixed point radius ($J=0$):
\beqn
\label{eq:NJP:app:RcEq}
R^{\rm eq}_c=\frac{\hat{P}_{\rm out} l_c}{\Delta} \ .
\eeqn
$R^{\rm eq}_c$ is unstable ($\left.dJ/dR\right|_{R_c}>0$) and is the nucleus radius at equilibrium ($k=h=0$): smaller droplets dissolve while larger droplets grow. 

When chemical reactions are switched on ($k,h>0$), \eq \eqref{eq:NJP:app:J} admits two fixed points, shown in \fig \ref{fig:NJP:appendix} for varying $k$. 
For small $R$ we can neglect the reaction term ($\propto k R^3$) and find the unstable fixed point, or nucleus radius,
\beqn
\label{eq:NJP:app:Rc}
R_c \simeq R_c^{\rm eq} \ ,
\eeqn
and for large $R$ we neglect the surface tension term ($\propto l_c/R$) and find the stable fixed point.
\beqn
\label{eq:NJP:app:Ru} 
R_u \simeq  \sqrt{ \frac{3D \Delta}{\hat{P}_{\rm in} k}} \ .
\eeqn
Additionally there exist a critical rate $k_c$ above which no fixed points exist and $J<0$ for all $R$ meaning that all droplets dissolve.

We will now examine these results in model A and C, seeking for the ratio $R_u/R_c$ for $k \simeq k_c$.

\subsection{Model C}
\label{sec:NJP:app:C}

In model C, $k=\alpha K_C$ and $h=\alpha H_C$ (\eq \eqref{eq:NJP:modelC}) so the ratio $k/h$ and $\Delta$ are constant. Therefore $R_u \propto \alpha^{-1/2}$ and we recover \eq \eqref{eq:NJP:scalingC}. Since this scaling is sub-linear we saw that it cannot explain SG formation and dissolution upon small variations of $\alpha$. However when $\alpha$ approaches $\alpha_c$ the separation between $R_c$ and $R_u$ becomes small so the above approximations cease to be valid and a strong response regime exists: small variations of $\alpha$ lead to strong variations of $R_u$ and $R_c$ ($\alpha\approx\alpha_c$ in  \fig \ref{fig:NJP:appendix}).

Qualitatively, it can be seen from \eq \eqref{eq:NJP:app:J} that since we omitted the term $\propto l_c/R$ in the determination of $R_u$ we have overestimated $R_u$, and since we neglected the term $\propto k R^3$ in the determination of $R_c$ we have underestimated $R_c$. Therefore the exact value of $R_u$ is bounded from above by $\sqrt{3D \Delta /( \hatPin k)}$ while the exact value of $R_c$ is bounded from below by $R_c^{\rm eq}$ (\eq \eqref{eq:NJP:app:RcEq}). The ratio $R_u/R_c$ in the strong response regime is therefore also bounded (see \fig \ref{fig:NJP:appendix}):
\beqn
\label{eq:NJP:appARuRc}
\frac{R_u}{R_c}< \left. \frac{\sqrt{3D \Delta /(\hatPin k)}}{R_c^{eq}} \right|_{k=K_C \alpha_c} \ .
\eeqn

At $\alpha=\alpha_c$ the two fixed points $R_c$ and $R_u$ intersect at the radius $R^*$ and since $R_c$ and $R_u$ are unstable and stable fixed points, respectively, we have
\beqn 
J(\alpha_c,R^*)&=&0 \\
\left. \frac{\dd J}{\dd R}\right|_{\alpha_c,R^*}&=&0 \ ,
\eeqn
and solving for $\alpha_c$ gives
\beqn
\label{eq:NJP:appAac}
\alpha_c= \frac{4D \Delta^3}{9l_c^2 \hat{P}_{\rm out}^2\hat{P}_{\rm in}K_C} \ .
\eeqn
 Using \eqs  \eqref{eq:NJP:modelC}, \eqref{eq:NJP:triangle} \eqref{eq:NJP:appARuRc} and \eqref{eq:NJP:appAac}, we then find
\beqn
\frac{R_u}{R_c} < \frac{3\sqrt{3}}{2} \ .
\eeqn
In other words, the size of the stable droplets and nuclei are of the same order. This shows independently of the system parameters that the strong response regime in model C cannot account for the switch-like response observed experimentally.

\subsection{Model A}
\label{sec:NJP:app:A}

We now concentrate on model A. Here the supersaturation $\Delta$ is no more constant since only the backward rate is constant (i.e. $k=\alpha K_A$, $h=H_A$):
\beqn
\label{eq:NJP:app:triangleA}
\Delta=\frac{\phi}{1+\alpha \frac{K_A}{H_A}} \ .
\eeqn
Therefore there exist a critical ATP concentration $\alpha_c$ above which $\Delta=0$ and all droplets dissolve ($R_u=0$, \eq \eqref{eq:NJP:app:Ru}). The expression of $\alpha_c$ is  found by solving $\Delta(\alpha_c)=0$:
\beqn
\label{eq:NJP:app:alpha*}
\alpha_c=\frac{\phi-\hatPo}{\hatPo} \frac{H_A}{K_A} \ .
\eeqn
We define  $\alpha_n \equiv 4 \alpha_c/3$ and $\alpha_n/2$ the ATP concentrations during  normal and stress condition, respectively, in agreement with the biological constraint that  ATP is depleted two-fold during stress conditions (Sec.~\ref{sec:NJP:experiments}). Moreover we assumed these concentrations to be equidistant from $\alpha_c$ for simplicity. During normal conditions  $\alpha=\alpha_n>\alpha_c$ so no droplets can exist  ($\Delta, R_u <0$). During stress condition  $\alpha=\alpha_n/2<\alpha_c$ and using \eqs \eqref{eq:NJP:app:Rc},\eqref{eq:NJP:app:Ru}, \eqref{eq:NJP:app:triangleA} and \eqref{eq:NJP:app:alpha*}, we find the size ratio $R_u/R_c$ during stress conditions:
\beq
\label{eq:NJP:appAFinal}
\frac{R_u( \alpha_n/2)}{R_c(\alpha_n/2)}=\frac{3}{4} \sqrt{ \frac{D}{l_c^2 H_A}\frac{(\hatPo)^2}{(\phi-\hatPo) \hatPin}  \left( \frac{\phi/\hatPo-2}{\phi/\hatPo+1}\right)^3 }
\eeq
Therefore we find that in model A and contrarily to model C, the size ratio between stable droplets and nuclei is function of the system parameters and can be arbitrarily large. This can potentially provide the switch-like response observed experimentally which we discuss quantitatively in  Sec.~\ref{sec:NJP:steadystate} using physiologically relevant parameters.

	\section{Determination of model parameters}
	\label{sec:NJP:param}
Focusing on  model $A$ with the backward reaction rate constant $H_A$ such that $H_A \ll \kappa$ regime, we now combine theory and SG experimental observations to determine the model parameters.
 As was done in \ref{sec:NJP:appendix}, let us denote $\alpha_n$ the ATP concentration during normal conditions, so that the ATP during stress condition is $\alpha_n/2$. For SG to form during stress and dissolve during normal conditions, we must have $\alpha_n/2 < \alpha_c < \alpha_n$, where $\alpha_c$ is the critical ATP concentration above which SG dissolve (\eq \eqref{eq:NJP:modelA:alpha_c} and \fig \ref{fig:NJP:modelA}(a)). For simplicity we assume that $\alpha_c$ is in the middle of the interval, i.e $\alpha_n-\alpha_c=\alpha_c-\alpha_n/2$. We thus find the expression of the ATP concentration during normal conditions from \eqs \eqref{eq:NJP:barP} and \eqref{eq:NJP:modelA:alpha_c}:
	\beqn
	\label{eq:NJP:param:alpha_n}
	\label{eq:NJP:constraint2}
	\alpha_n=\frac{4 H_A}{3 K_A} \left( \frac{\phi}{\hatPo}-1\right)  \ .
	\eeqn

	\subsection{Fluorescence recovery after photobleaching (FRAP)}
	
	We employ fluorescence recovery after photobleaching (FRAP) experiments \cite{axelrod_biophys76} to estimate the diffusion coefficient of SG proteins. In these experiments a specific protein is tagged with a fluorescent marker, so that the local concentration of protein-marker is correlated with the local intensity of the fluorescence emission. A photobleaching laser beam is used to eliminate the fluorescence in a small region. Over time, the intensity in this region recovers as the non-fluorescent protein-markers are gradually replaced by fluorescent protein-markers, due to diffusion or other transport mechanisms. Assuming free diffusion of the protein-marker, the diffusion coefficient is given by  $D \approx w^2 / (4 t_{1/2})$ with $t_{1/2}$ the half-time of fluorescence intensity recovery and $w$ the radius of the bleached region \cite{axelrod_biophys76}.
	
	We now consider FRAP results on TIA proteins. TIA proteins are key SG constituents, triggering SG formation when over-expressed, and inhibiting SG assembly when depleted \cite{gilks_mboc04}.
	Moreover TIA proteins harbor an intrinsically disordered domain, a class of protein domains that have raised recent interest for its ability to form phase-separated liquid droplets \cite{uversky2016}. Hence, TIA protein is a prime candidate to explain SG formation based on phase separation.
	
	FRAP experiments on TIA proteins show that the intensity of a bleached cytoplasmic region of radius $\sim 2 \mu m$ recovers with a half-time $t_{1/2} \simeq 2 s$ \cite{bley_17}. This leads to the protein diffusion coefficient
	\beqn
	\label{eq:NJP:constraintD}
	D \approx 0.5 \rm ~ \mu m^2 s^{-1} \ .
	\eeqn
	
	Interestingly, the half-time recovery in experiments where a SG is bleached are roughly similar \cite{bley_17}. This indicates that SG proteins are highly mobile inside SG and supports our approximation of identical protein diffusion coefficients inside and outside SG (\eqs \eqref{eq:NJP:P} and \eqref{eq:NJP:SS}). 
	We note that such experiments are more complicated to interpret than cytoplasmic FRAP, because of the added parameters influencing the recovery time, such as the concentration ratio between SG and cytoplasm.
	
	\subsection{Robustness of the model against concentration fluctuations}
	\label{sec:NJP:robust}

	The global protein concentration in a cell is subject to fluctuations \cite{kaern_nature05}. This can result in an increase of the overall concentration $\phi$ of SG constituent molecules, and potentially trigger undesired SG formation during non-stress conditions. This is due to the subsequent increase of the overall concentration $\bar P$ of phase-separating states (\eq \eqref{eq:NJP:barP}), driving the system inside the phase-separating region ($\bar P>\hatPo$, \fig \ref{fig:NJP:fig1}). Let us define $\phi'$ the concentration of SG constituents such that, during normal conditions ($\alpha=\alpha_n$), $\bar P$ is identical to its value during stress conditions ($\alpha=\alpha_n/2$), thereby leading to undesired SG formation. In other words $\phi'$ is the solution of $\bar P(\alpha_n,\phi')=\bar P(\alpha_n/2,\phi)$ (\eq \eqref{eq:NJP:barP} and \eqref{eq:NJP:modelA}). The ratio $\phi'/\phi$ should be large enough to prevent SG formation in normal conditions. From the definition of $\phi'$ and \eqs \eqref{eq:NJP:barP} and \eqref{eq:NJP:param:alpha_n} we find:
	\beqn
	\label{eq:NJP:phiphip}
	\frac{\phi'}{\phi} = \frac{4 \phi/\hatPo-1}{2 \phi/\hatPo+1} \ .
	\eeqn
	The ratio $\phi'/\phi$ is an increasing function of $\phi/\hatPo$, and reaches a plateau for $\phi/\hatPo \gg 1$ (continuous purple line in \fig \ref{fig:NJP:param}). Note that if $\phi/\hatPo<1$ the system does not phase separate, since $\bar P < \hatPo$ in this case (\eq \eqref{eq:NJP:barP}, \fig \ref{fig:NJP:fig1}). If $\phi \gtrsim \hatPo$ then $\phi' \gtrsim \phi$ and even small fluctuations of $\phi$ cause unwanted SG formation. This scenario is not physiological and we discard it. If on the contrary $\phi/\hatPo \gg 1$ then $\phi' \approx 2\phi$. In this case, a two-fold increase of $\phi$ is necessary to bring $\bar P$ to its stress value. This scenario is therefore robust against concentration fluctuations. This prediction is consistent with experimental observations showing that a three-fold increase of the concentration of a given SG protein triggers SG formation \cite{reineke_mboc12}. Therefore we must impose:
	\beqn
	\label{eq:NJP:constraint1}
	\frac{\phi}{\hatPo} \gg 1 \ .
	\eeqn
	We will see in \ref{sec:NJP:ratio} that $\phi/\hatPo$ is also maximally bounded.

	\subsection{Partitioning coefficient}
	\label{sec:NJP:partition}
	
	We use experimental results on TIA partitioning coefficient, i.e. the concentration ratio between SG and cytoplasm, to constrain further the model parameters. The combined concentrations of $P+S$ inside and outside a SG of radius $R$ are given by $\hatPin + S(R)$ and $\hatPo + S(R)$, respectively \cite{wurtz_prl18}, and we recall that $S(R)$ is the concentrations of states $S$ at the droplet interface (\eq \eqref{eq:NJP:S}). Note that the combined concentration of $P+S$ are homogeneous inside and outside, unlike the individual concentrations of $P$ and $S$ (\eq \eqref{eq:NJP:profile} and \fig \ref{fig:NJP:profile}). The partitioning coefficient $a$ is therefore given by
	\beqn
	\label{eq:NJP:param:a1}
	a=\frac{\hatPin + S(R)}{\hatPo + S(R)} \ .
	\eeqn
	The value of $S(R)$ depends on the model parameters and the number of droplets in the system, and is bounded as follows \cite{wurtz_prl18}:
	\beqn
	\frac{K_A \alpha_n \phi }{K_A \alpha_n + H_A} < S(R) < \phi-\hatPo \ .
	\eeqn
	This result with \eqs \eqref{eq:NJP:constraint2} and \eqref{eq:NJP:constraint1} leads to $S(R) \approx \phi$ and  $a \simeq (\hatPin + \phi)/\phi$. Experimental observations on the partitioning coefficient $a$ of TIA proteins is of order 10 \cite{bley_17}, leading to the following constraint on the concentration parameters:
	\beqn
	\label{eq:NJP:constraint4}
	\hatPin \approx 10 ~  \phi \ .
	\eeqn

	\subsection{Size ratio between SG nuclei and mature SG}
	\label{sec:NJP:ratio}
	
	\begin{figure}
		\centering
		\includegraphics[scale=0.9]{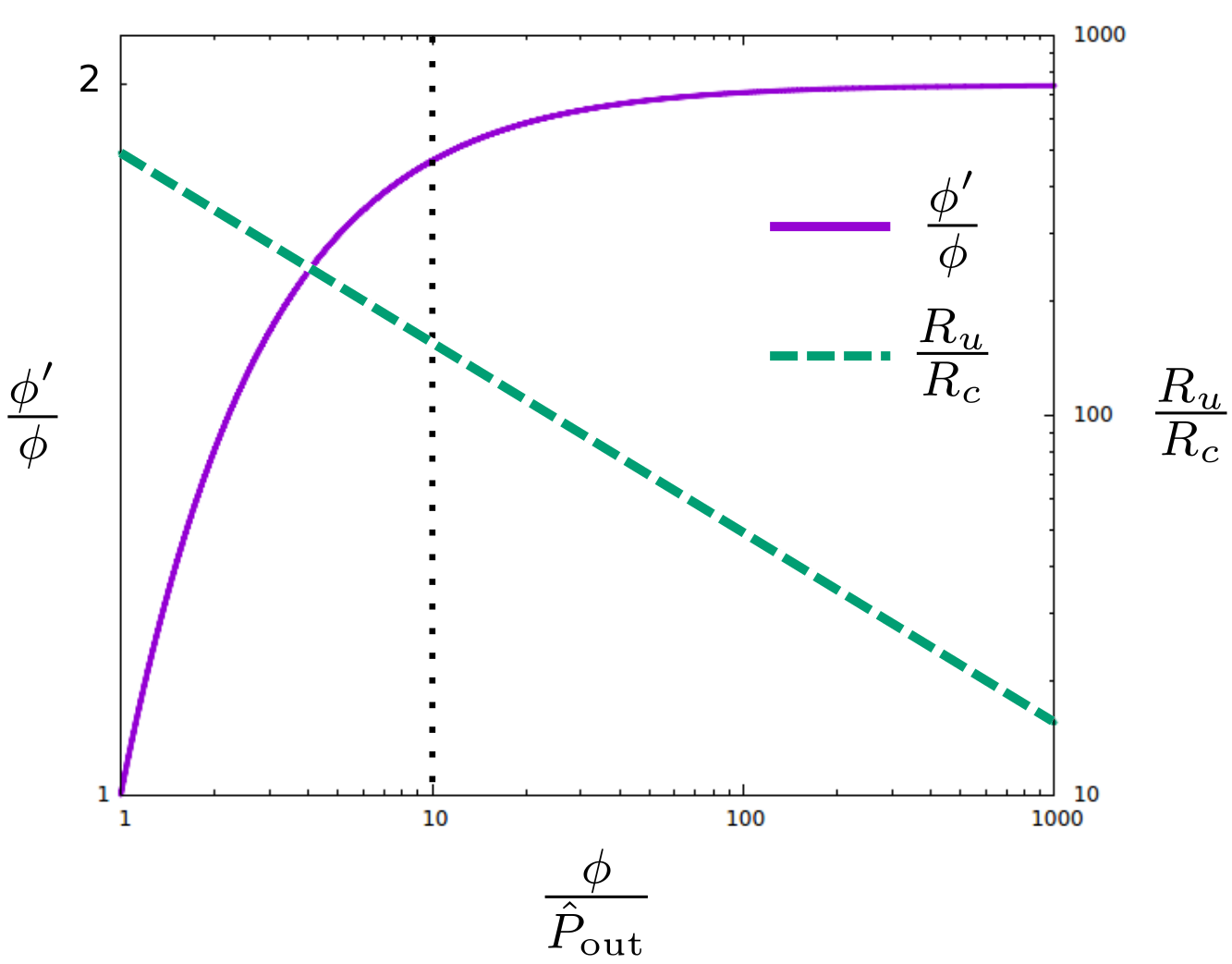}
		\caption{
			\textit
			{ 
					The robustness of the model against concentration fluctuations (quantified by $\phi'/\phi$ , purple curve, \eq \eqref{eq:NJP:phiphip}), and the radius ratio $r$ between mature SG and SG nuclei (which is upper-bounded by $R_u/R_c$, green dashed line, \eq \eqref{eq:NJP:r}, arbitrary units), are functions of the concentration ratio $\phi/\hatPo$. The choice of $\phi/\hatPo \approx 10$ (vertical dotted line) ensures both model robustness and high ratio $R_u/R_c$.
			}
			\label{fig:NJP:param}		
		}  
	\end{figure}
	
	From experimental observations we know that mature SG are much larger than the SG nuclei  $R_c$ (\eq \eqref{eq:NJP:theory:RcRlRu}) (Sec. \ref{sec:NJP:experiments}). Since $R_u$ is the radius of the largest possible SG (\eq \eqref{eq:NJP:theory:RcRlRu}), the radius ratio $r$ between mature SG and nuclei is upper-bounded as follow:
	\beqn
	r<\frac{R_u}{R_c} \ .
	\eeqn
	Using the expression of $R_u/R_c$ obtained in \ref{sec:NJP:appendix} (\eq \eqref{eq:NJP:appAFinal}), and imposing the constraints from \eqs  \eqref{eq:NJP:param:alpha_n}, \eqref{eq:NJP:constraint1}, and \eqref{eq:NJP:constraint4} on the ratio $r$, we find
	\beqn
	\label{eq:NJP:r}
	r<\sqrt{ \frac{3D \hatPo}{40 K_A \alpha_n l_c^2 \phi}} \ . 
	\eeqn

	In \ref{sec:NJP:robust} we have shown that $\phi / \hatPo \gg 1$ must be true to ensure robustness of the model against concentration fluctuations. Here we see that $\phi/\hatPo$ cannot be arbitrarily large since $r$ would approach 0, as shown in \fig \ref{fig:NJP:param} (green dashed line). In order to ensure both model robustness against concentration fluctuations and large radius ratio $r$ between mature SG and SG nuclei, we choose the following constraint (dotted vertical line in \fig \ref{fig:NJP:param})
	\beqn
	\label{eq:NJP:constraint6}
	\frac{\phi}{\hatPo }= 10 \ .
	\eeqn
	
	Note that since $r$ is also a function of other model parameters ($D,K_A,\alpha_n,lc$), the value of $\phi/\hatPo$ could in principle be much larger than $10$, as long as the other parameters are adequate. We will see however that using physiologically relevant parameters determined in this section, the predicted ratio $r$ is consistent with experimental observations. Therefore taking $\phi/\hatPo$ much larger than $10$ would lead to an unrealistically small ratio $r$. This supports our choice in \eq \eqref{eq:NJP:constraint6}.
	
	\subsection{SG formation and dissolution time scales}
	\label{sec:NJP:reactionTimes}
	
	In our model SG form and dissolve in response to changes of the overall concentration $\bar P$ of the phase-separating states. To be consistent with experimental observations, the predicted time scales at which $\bar P$ varies 
	must be shorter than the SG formation and dissolution time scales. This imposes constraints on the reaction rate constants $K_A$ and $H_A$ (\eq \eqref{eq:NJP:dPdt}). 
	Let us assume for simplicity that the variation of $\bar P$ occurs at constant ATP concentration $\alpha$. With $\bar P$ being the only time-dependent variable in \eq \eqref{eq:NJP:dPdt}, we can solve and we find
	\beqn
	\label{eq:NJP:P(t)}
	{\bar P}=c_1 + c_2 e^{-( K_A \alpha  + H_A )t}  \ ,
	\eeqn
	with $c_i$ functions of the parameters $K_A, H_A, \phi, \alpha$. The time scale at which $\bar P$ reaches chemical equilibrium is given by the inverse of $K_A \alpha  + H_A$. Since the ATP concentration $\alpha$ varies only by a factor 2 between stress and normal conditions (Sec. \ref{sec:NJP:experiments}), the increase and decrease of $\bar P$ occur on a similar time scale $ \tau \approx 1/(K_A \alpha_n+H_A)$, with $\alpha_n$ the ATP concentration during normal conditions (\eq \eqref{eq:NJP:param:alpha_n}). This is consistent with the experimental observations that SG formation and dissolution times are within the same order of magnitude. 
	
	As a rough estimate, $\tau$ must be smaller than the SG formation/dissolution time. 
	Taking this time scale to be 15 minutes we get $H_A + K_A \alpha_n > 10^{-3} \rm~ s^{-1}$. Since $K_A \alpha_n \gg H_A$ (from \eqs \eqref{eq:NJP:param:alpha_n} and \eqref{eq:NJP:constraint1}) we obtain $K_A \alpha_n > 10^{-3} \rm~ s^{-1}$. Taking an unnecessary large value of $K_A \alpha_n$ would lead to an ATP consumption that is unnecessary high. We therefore choose a conservative value of $K_A \alpha_n$, just large enough to ensure that the time scale of $\bar P$ variation is fast enough:
	\beqn
	\label{eq:NJP:constraint7}
	K_A \alpha_n \approx 2 \times 10^{-3} \rm s^{-1} \ .
	\eeqn
 We note that this rate constant is consistent with typical rates of protein phosphorylation  \cite{rose_jbc76,aoki_scientificreport13}, which is the prototypical ATP-dependent reaction for protein post-transcriptional modification \cite{ardito_ijmm17}. Using this result with \eqs \eqref{eq:NJP:param:alpha_n} and \eqref{eq:NJP:constraint6} we find the backward reaction rate constant:
	\beqn
	\label{eq:NJP:constraint9}
	H_A\approx 1.5 \times 10^{-4} \rm s^{-1} \ .
	\eeqn

	\subsection{Nucleus radius}
	
	\label{sec:NJP:Rc}
	
	The capillary length $l_c$ can be estimated from the expected SG nucleus radius $R_c$. The expression for $R_c$ is given by \eq \eqref{eq:NJP:theory:RcRlRu}:
	\beqn
	R_c = \frac{\hatPo l_c}{H_A \phi/(H_A+0.5 K_A \alpha_n)-\hatPo} \ .
	\eeqn
	where we used the ATP concentration during stress: $\alpha_n/2$. Enforcing the constraints from \eq \eqref{eq:NJP:param:alpha_n} and \eqref{eq:NJP:constraint1} we find
	\beqn
	R_c \approx 3.3 l_c \ .
	\eeqn
	
	The SG nucleus radius $R_c$ is necessarily smaller than the radius of the smallest detectable SG ($\sim 100 \rm nm$). On the other hand, there exist indications that SG are nucleated \textit{via} heterogeneous nucleation involving RNA as seeding platforms \cite{zhang_cell15, altmeyer_nature15}. This implies that SG must reach a critical size before they become thermodynamically stable. Therefore SG nuclei are larger than the dimension of a single protein, which is of the order of $5 \rm nm$. We choose an intermediate value, $R_c=50 \rm nm$, leading to an estimation of the capillary length:
	\beqn
	\label{eq:NJP:constraintlc}
	l_c\approx 15~  \rm nm \ .
	\eeqn
	
	\subsection{Summary}

	We have used theory and SG experimental results to determine physiologically relevant model parameters, or parameter constraints, summarized by \eqs \eqref{eq:NJP:constraintD}, \eqref{eq:NJP:constraint4}, \eqref{eq:NJP:constraint6}, \eqref{eq:NJP:constraint7}, \eqref{eq:NJP:constraint9} and \eqref{eq:NJP:constraintlc}. Note that the concentration parameters $\phi,~ \hatPo,~ \hatPin$, the ATP concentration $\alpha_n$, and the forward reaction constant $K_A$ have not been explicitly expressed. Yet, the radius $R_c$ of the SG nucleus, and the radii $R_l$ and $R_u$ that bound the stable region (\eq \eqref{eq:NJP:theory:RcRlRu}) are now explicitly determined from these relations.

	\section{Equivalence between equilibrium and non-equilibrium regimes for the early droplet growth kinetics}
	\label{sec:NJP:appendix:equivalence}
	
	We consider the small droplet regime ($R\ll\xi$, \eq \eqref{eq:NJP:xi}, Sec. \ref{sec:NJP:theory}) and show that, far from the steady-state, the droplet growth dynamics is equivalent to that in an equilibrium system, i.e. without chemical reactions but with identical amounts of $P$ and $S$.
	
	\subsection{Small droplet number density}
	\label{sec:NJP:appendix:equivalence:small}
	
	 Let us concentrate first on the situation where the droplet number density is small, so that the inter-droplet distance is much larger than the gradient length scale $\xi$. The term $U_{\rm out}^{(1)}$ in \eq \eqref{eq:NJP:profile} must be zero to prevent diverging concentration $P_{\rm out}(r)$ far from droplets. Applying the boundary condition \eq \eqref{eq:NJP:gt}, we obtain the concentration profile of $P$ in the cytoplasm:
	\beqn
	\label{eq:NJP:Pout1}
	P_{\rm out}(r) = P_{\infty} - \frac{R}{r} \left(\Delta - \frac{ \hatPo l_c}{R} \right) e^{-(r-R)/\xi} \ ,
	\eeqn
	where $P_\infty$ is the cytoplasmic concentration of $P$ far from droplets, or far-field concentration, and $\Delta \equiv P_{\infty} - \hatPo $ is the supersaturation concentration. The influx of $P$ material from the cytoplasm at the droplet interface ($4 \pi D R^2  \dd P_{\rm out} / \dd r |_{r=R}$) is 
	\beqn
	\label{eq:NJP:Jin}
	4 \pi D R \left(\Delta - \frac{ \hatPo l_c}{R} \right)  \ ,
	\eeqn
	where we have used the small droplet condition ($R\ll\xi$). In this small droplet regime the concentration inside droplet is homogeneous ($P_{\rm in}(r)=\hatPin$). Therefore the depletion rate of $P$ material inside droplets by the chemical conversion $P \rightarrow^k S$ is given by 
	\beqn
	\label{eq:NJP:depletion}
	\frac{4 \pi R^3}{3} k_i(\alpha) \hat{P}_{\rm in} \ .
	\eeqn
	We can now write the total influx of $P$ at the droplet interface:
	\beqn
	\label{eq:NJP:theory:J}
	J= 4 \pi D R \left(\Delta - \frac{ \hat{P}_{\rm out}l_c}{R}\right)-\frac{4 \pi R^3}{3}k_i(\alpha) \hat{P}_{\rm in} \ ,
	\eeqn
	where the first term accounts for the influx from the cytoplasm (\eq \eqref{eq:NJP:Jin}) and the second term accounts for the chemical conversion of $P$ into $S$ inside droplets (\eq \eqref{eq:NJP:depletion}). At small droplet radius $R$, the reaction term ($\propto k_i$) in \eq \eqref{eq:NJP:theory:J} is negligible compared to the diffusive term. The droplet growth dynamics can therefore be mapped onto the growth dynamics in an equilibrium system \cite{lifshitz_jpcs61}, where the supersaturation $\Delta \equiv P_\infty - \hatPo$ is controlled by the chemical reaction rates. At small enough droplet number density, since the largest possible droplet radius, $R_u$ (\eq \eqref{eq:NJP:theory:RcRlRu}), is independent of the system size, the presence of droplets does not affect the far-field concentration $P_\infty$. Therefore $P_\infty$ is equal to $\bar P$ (\eq \eqref{eq:NJP:barP}), the overall concentration of $P$ in the system. The value of the supersaturation $\Delta$ is therefore equal to that in equilibrium condition, in the diffusion limited growth regime ($\Delta=\bar P - \hatPo$) \cite{pitaevskii_b81}. 
	
	\subsection{Small droplets number density}
	
	If the droplet number density is large so that the inter-droplet distance is small compared to $\xi$, we can neglect the effect of the chemical reactions on the profiles and we recover the equilibrium profile \cite{lifshitz_jpcs61}:
	\beqn
	\label{eq:NJP:Pout2}
	P_{\rm out}(r) = P_{\infty} - \frac{R}{r} \left(\Delta - \frac{\hatPo l_c}{R} \right) \ .
	\eeqn
	This result can be verified by expanding \eq \eqref{eq:NJP:profile} for the small parameters $R/\xi$ and $L/\xi$ where $L$ is the inter-droplet distance, and by imposing the boundary conditions \eqs \eqref{eq:interfacePin}-\eqref{eq:NJP:boundary2}.
	Similarly as in \ref{sec:NJP:appendix:equivalence:small} the depletion rate of $P$ inside droplets is given by \eq \eqref{eq:NJP:depletion} and therefore the total influx of $P$ at the droplet interface is given by \eq \eqref{eq:NJP:theory:J}. Again, for small droplet radius $R$ the reaction term in \ref{eq:NJP:theory:J} is negligible and the droplet growth dynamics can be mapped onto an equilibrium growth.	Since the cytoplasmic profile is identical to that in an equilibrium system (\eq \eqref{eq:NJP:Pout2}), the same must be true for $\Delta$. Therefore the droplet growth dynamics is identical to that at equilibrium condition, and if the number of droplet is large enough, $\Delta$ is small and the system undergoes Ostwald ripening \cite{pitaevskii_b81}. 
	
	These equivalences between equilibrium and non-equilibrium regimes allow us to estimate SG growth dynamics far from the steady-state (Sec. \ref{sec:NJP:dynamics}).

\section*{References}

\bibliographystyle{unsrt}

\providecommand{\newblock}{}

%
%
%

\end{document}